\documentclass{article}

\usepackage{microtype}
\usepackage{graphicx}
\usepackage{subcaption}
\usepackage{booktabs} 

\usepackage{hyperref}

\usepackage[accepted]{icml2026}

\usepackage{amsmath}
\usepackage{amssymb}
\usepackage{mathtools}
\usepackage{amsthm}
\usepackage{multirow}
\usepackage{makecell}
\usepackage{enumitem}
\usepackage{tabularx}

\usepackage{eso-pic}
\usepackage{xcolor}

\usepackage[capitalize,noabbrev]{cleveref}

\theoremstyle{plain}

\theoremstyle{definition}

\theoremstyle{remark}

\usepackage[disable,textsize=tiny]{todonotes}

\icmltitlerunning{MECAT: A Benchmark for Fine-Grained Audio Understanding}

\begin{document}

\twocolumn[
  \icmltitle{MECAT: A Multi-Experts Constructed Benchmark for \\
  Fine-Grained Audio Understanding Tasks}



  \icmlsetsymbol{equal}{*}

  \begin{icmlauthorlist}
    \icmlauthor{Yadong Niu}{equal,xiaomi}
    \icmlauthor{Tianzi Wang}{equal,xiaomi,cuhk}
    \icmlauthor{Heinrich Dinkel}{xiaomi}
    \icmlauthor{Xingwei Sun}{xiaomi}
    \icmlauthor{Jiahao Zhou}{xiaomi}
    \icmlauthor{Gang Li}{xiaomi}
    \icmlauthor{Jizhong Liu}{xiaomi}
    \icmlauthor{Xunying Liu}{cuhk}
    \icmlauthor{Junbo Zhang}{xiaomi}
    \icmlauthor{Jian Luan}{xiaomi}
  \end{icmlauthorlist}

  \icmlaffiliation{xiaomi}{MiLM Plus, Xiaomi Inc, Beijing, China}
  \icmlaffiliation{cuhk}{The Chinese University of Hong Kong, Hong Kong, China}

  \icmlcorrespondingauthor{Yadong Niu}{niuyadong@xiaomi.com}
  \icmlcorrespondingauthor{Heinrich Dinkel}{dinkelheinrich@xiaomi.com}
  \icmlcorrespondingauthor{Tianzi Wang}{twang@se.cuhk.edu.hk}

  \icmlkeywords{Machine Learning, ICML}

  \vskip 0.3in
]



\printAffiliationsAndNotice{\icmlEqualContribution}

\begin{abstract}
While large audio-language models have advanced open-ended audio understanding, they still fall short of nuanced human-level comprehension. This gap persists largely because current benchmarks, limited by data annotations and evaluation metrics, fail to reliably distinguish between generic and highly detailed model outputs. To this end, this work introduces MECAT, a \textbf{M}ulti-\textbf{E}xpert \textbf{C}onstructed Benchmark for Fine-Grained \textbf{A}udio Understanding \textbf{T}asks. Generated via a pipeline that integrates analysis from specialized expert models with Chain-of-Thought large language model reasoning, MECAT provides multi-perspective, fine-grained captions and open-set question-answering pairs. The benchmark is complemented by a novel metric: DATE (\textbf{D}iscriminative-\textbf{E}nhanced Audio \textbf{T}ext \textbf{E}valuation). This metric penalizes generic terms and rewards detailed descriptions by combining single-sample semantic similarity with cross-sample discriminability. A comprehensive evaluation of state-of-the-art audio models is also presented, providing new insights into their current capabilities and limitations. The code and data are publicly available at \url{https://github.com/xiaomi-research/mecat}.
\end{abstract}

\section{Introduction}

The human auditory system is highly effective at processing complex acoustic scenes. It can distinguish subtle variations in sound, such as telling a dog's playful bark from a defensive growl~\citep{plack2023sense}, and isolate target speech from noisy backgrounds, an ability known as the cocktail party effect.
A central goal of machine hearing is to replicate this auditory intelligence to interpret raw audio signals as semantically rich perception~\citep{lyon2017human}. Early works in machine hearing focused on closed-ended tasks such as sound event classification and automatic speech recognition. Large language models (LLM) have spurred the development of large audio-language models (LALMs), which have driven a shift towards more general open-ended tasks like audio captioning and audio question answering~\citep{chu2023qwen,du2023lauragpt,hu2024wavllm,shu2023llasm,wang2023blsp,tang2023salmonn,rubenstein2023audiopalm,chen2023x,huang2024audiogpt}.

Despite these advances, current LALMs still fall short of achieving the comprehensive understanding that characterizes human hearing~\citep{sakshi2025mmau}. This work argues that despite ongoing improvements in model architectures and data, a crucial and often-overlooked bottleneck is the existing evaluation benchmark. 

The first challenge lies with data annotations, which suffers from several limitations. To begin with, the annotations in current benchmarks are often overly simplistic, consisting of event-level captions that lack detail \cite{mei2024wavcaps, kim2019audiocaps, drossos2020clotho} or question-answering tasks confined to close-ended formats \cite{lipping2022clotho, wang2025audiobench, sakshi2025mmau}. Furthermore, they typically adopt a monolithic perspective, providing a single, global description that fails to account for the selective nature of human hearing. Compounding these issues is a ``one-to-many'' data redundancy problem, where the same audio clips, often from AudioSet~\cite{gemmeke2017audio}, are reused across multiple benchmarks, limiting the assessment of model generalization.

The second challenge is rooted in evaluation metrics. Traditional lexical-matching metrics, on the one hand, penalizes semantically correct but lexically different descriptions. Embedding-based metrics, on the other hand, better align with human perception, they often fail to distinguish between generic, vague captions and highly detailed, accurate ones. Even the more recent LLM-as-judge method, while demonstrating strong discriminative ability, is often hindered by practical constraints such as high costs and slow inference speeds, as well as its high dependency on model selection and prompt design.

Thus, current benchmarks inadequately evaluate audio understanding, as they often reward generic captions (e.g., A dog is barking and people are talking) for distinct scenarios (e.g., an excited bark in a park vs. a defensive bark during an argument). 
This limits their ability to differentiate between models with true perceptual accuracy and those producing vague outputs.

To this end, we introduce {MECAT}, a Multi-Expert Constructed Benchmark for Fine-Grained Audio Understanding Tasks. By integrating analysis from a series of specialized audio-related experts models, including content-specific models (e.g., for speech, music, and sound events) and content-unrelated models (e.g., for audio quality, reverberation and intensity), followed by Chain-of-Thought (CoT) enhanced LLM reasoning~\citep{guo2025deepseek}, MECAT provide fine-grained captions alongside open-set question-answering pairs. The captions primarily focus on providing a comprehensive, multi-perspective description of the acoustic scene, while the QA pairs are designed to probe for specific details and higher-level contextual reasoning that descriptive tasks alone cannot fully assess. Furthermore, we introduce a novel metric \textit{DATE} (Discriminative-Enhanced Audio Text Evaluation), which is designed to better quantify the detail and accuracy of model's response. 
It uniquely combines a weighted single-sample semantic similarity that penalizes generic terms while emphasizing discriminative phrases, and a cross-sample discriminability score that explicitly rewards the model's responses for exceeding general descriptions. This design enables DATE to robustly distinguish between superficial and context-rich model outputs.

\paragraph{Conflict of Interest Disclosure.} Several authors are employed by Xiaomi Inc., the developer of the MiMo-Audio model evaluated in this paper.

\section{Related works}
\subsection{Audio Captioning Benchmark}

Audio captioning benchmarks have been pivotal in advancing audio understanding  works~\citep{wu2019audiocaption_listen_and_tell,kim2019audiocaps,drossos2020clotho,yuan2025sound,manco2023song,liu2024enhancing, liu2024leveraging}. 
Early dataset like AudioCaps~\citep{kim2019audiocaps} and Clotho~\citep{drossos2020clotho} primarily relied on manual annotation, where human annotators provide one or more captions for each audio clip. While foundational, these benchmarks face a critical limitation: the coarse-grained nature of their annotations. During the annotation process, human annotators often produce generic, events-level descriptions rather than capturing the nuanced acoustic details of a scene. This results in a gold standard that lacks the specificity needed to evaluate fine-grained understanding.

While newer methods using LLMs for automatic labeling, such as in AutoACD~\citep{sun2024auto} and LPMusicCaps~\citep{doh2023lp}, have improved scalability, they did not solve the granularity problem. 
Caption quality suffers from coarse input metadata like titles and tags, perpetuating generic descriptions.

\subsection{Audio Question-Answering Benchmark}

Audio Question Answering (QA) presents a more targeted evaluation of a model's audio understanding abilities~\citep{lipping2022clotho,wang2025audiobench,li2022learning,sakshi2025mmau, ma2025mmar}. Datasets like ClothoAQA~\citep{lipping2022clotho} and MusicAVQA~\citep{li2022learning} have been developed with manually crafted question-answer pairs. However, similar to captioning benchmarks, they suffer from limitations that hinder the assessment of detailed understanding.

The main issue is their reliance on close-ended answer formats designed for easier automatic scoring. For example, many questions in ClothoAQA are limited to "yes/no" answers~\citep{lipping2022clotho}, while other benchmarks like MMAU~\citep{sakshi2025mmau} utilize a multiple-choice format. While convenient for evaluation, these formats prevent the assessment of a model's ability to generate detailed, descriptive answers and may encourage models to learn shallow pattern matching rather than deep understanding.

\subsection{Evaluation Metrics for Audio Caption and QA}
The evaluation of open-ended audio caption and QA is critically dependent on the choice of metric. However, existing metrics fail to adequately assess the fine-grained descriptive capabilities of modern generative models. 

Traditional metrics, such as BLEU~\citep{papineni2002bleu}, CIDEr~\citep{vedantam2015cider}, and SPICE~\citep{anderson2016spice}, operate by measuring lexical overlap with reference texts. This reliance on n-gram matching unfairly penalizes novel yet accurate descriptions that do not share the exact wording of the references. 

\begin{figure*}[h]
  \centering
  \includegraphics[width=0.95\linewidth]{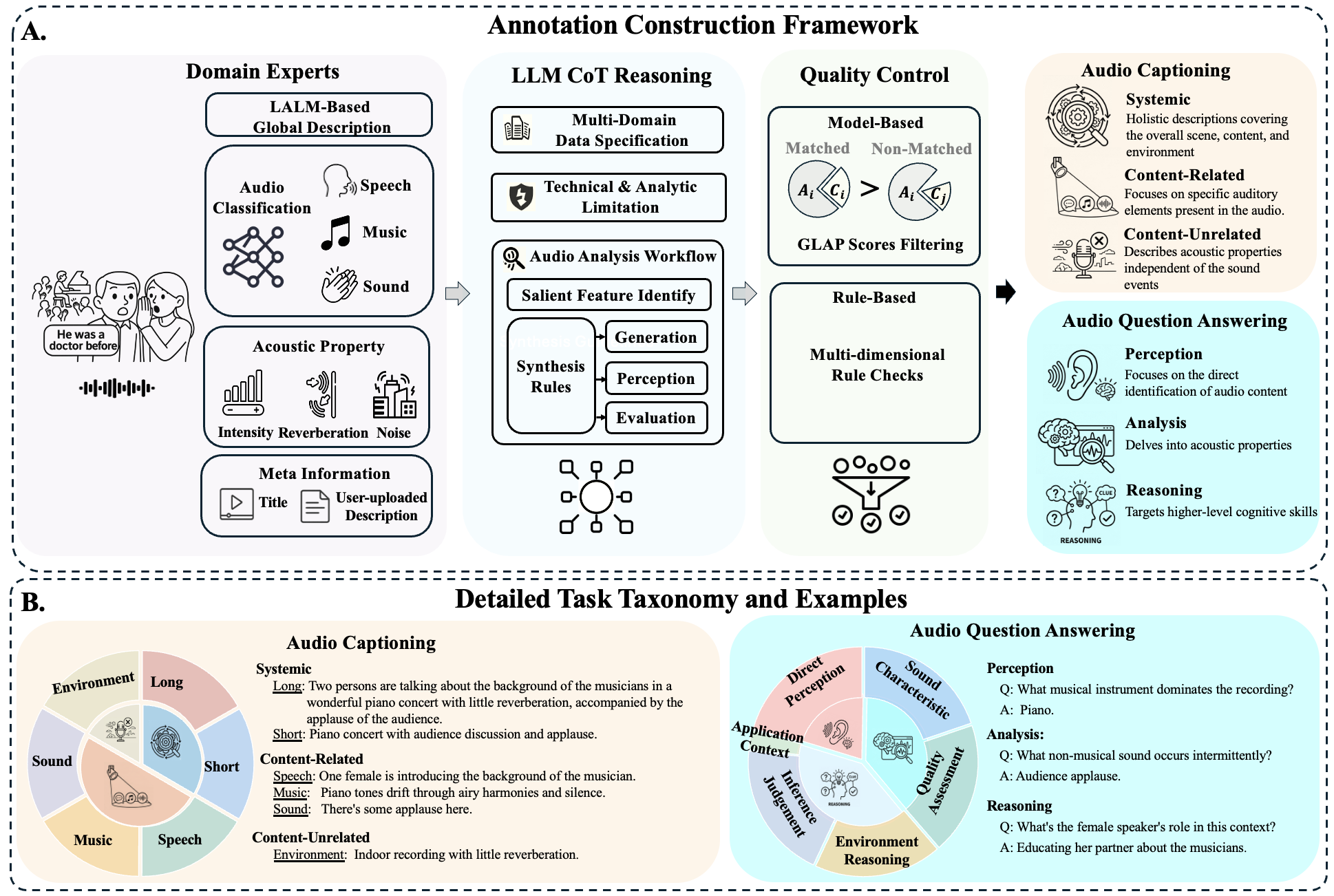}
  \caption{Overview of the MECAT Benchmark. (A) The proposed annotation construction pipeline. (B) Detailed task taxonomy and representative examples for Audio Captioning and Question Answering, showcasing the diversity of the dataset.}
  \label{fig:labeling-main}
  \vspace{-3mm}
\end{figure*}
To overcome the limitations of lexical matching, embedding similarity-based metrics were introduced. 
Approaches like FENSE~\citep{zhou2022fense} were specifically designed for audio captioning. 
However, our experiments found that they still struggle to effectively distinguish between a generic, vague response and a highly detailed and accurate one. 
More recently, LLM-as-judge has been adopted for evaluating open-ended generation~\citep{wang2025audiobench,zheng2023judging}. 
These methods show strong correlation with human judgment and, as our experiments confirmed, possess a high sensitivity to response specificity.
However, LLM-as-judge suffer from practical limitations, such as high computational costs, slow evaluation speeds, and a strong dependency on the choice of the LLM and the design of the prompt~\citep{lee2024fleur,zheng2023judging}.

\section{MECAT Benchmark Overview}
As illustrated in~\Cref{fig:labeling-main}, MECAT is a comprehensive benchmark  for fine-grained audio understanding, distinguished by its unique data sources, broad domain coverage, and two core evaluation tasks: MECAT-Caption and MECAT-QA.

\subsection{Dataset Description}
\label{sec:datasource}

\begin{table*}[ht]
\centering
\caption{Comparison of MECAT with Recent General Sound Evaluation Benchmark Datasets. ~$\dagger$ MP-LLM: Multiple Experts Models and LLM; $\ddagger$ Multi-Domain: This includes speech, music and sound-events ($\diamond$ denotes that domain were not elaborated in detail); $\S$ Extended Multi-Domain: This includes speech, music, sound-events, combinations thereof, and silence.}
\label{tab:benchmark_comparison}
\resizebox{\linewidth}{!}{%
\begin{tabular}{llllll}
\toprule
Task & Labeling & Dataset & Test Size & Domain & Source \\
\midrule
\multirow{8}{*}{{Caption}} & 
\multirow{3}{*}{\makecell[l]{Manual}} & 
AudioCaps ~\citep{kim2019audiocaps} & $\sim$1.0k & Multi-Domain\textsuperscript{$\ddagger$,$\diamond$} & AudioSet \\
& & Clotho \citep{drossos2020clotho} & $\sim$1.0k & Multi-Domain\textsuperscript{$\ddagger$,$\diamond$} & Clotho \\
& & SongDescriber ~\citep{manco2023song} & 0.7k & Music & MTG-Jamendo \\
\cmidrule(lr){2-6} 
& \multirow{4}{*}{\makecell[l]{LLM}} & 
AudioCaps-Enhanced ~\citep{yuan2025sound} & 0.9k & Multi-Domain\textsuperscript{$\ddagger$,$\diamond$} & AudioSet \\
& & AutoACD ~\citep{sun2024auto} & 1.0k & Multi-Domain\textsuperscript{$\ddagger$,$\diamond$} & AudioSet \\
& & LPMusicCaps-MSD ~\citep{doh2023lp} & 35k & Music & Song Dataset \\
& & LPMusicCaps-MTT ~\citep{doh2023lp} & 5k & Music & MagnaTagATune \\
\cmidrule(lr){2-6} 
& \multirow{1}{*}{\makecell[l]{MP-LLM\textsuperscript{$\dagger$}}} & 
MECAT-Caption (Ours) & 20k & Extended~Multi-Domain\textsuperscript{$\S$} & ACAV100M \\
\midrule
\multirow{8}{*}{{QA}} & 
\multirow{5}{*}{\makecell[l]{Manual}} & 
ClothoAQA ~\citep{lipping2022clotho} & 2k & Multi-Domain\textsuperscript{$\ddagger$,$\diamond$} & Clotho \\
& & WavCaps-QA ~\citep{wang2025audiobench} & 0.3k & Multi-Domain\textsuperscript{$\ddagger$,$\diamond$} & AudioSet and 2 others \\
& & MusicAVQA ~\citep{li2022learning} & 6k & Music & YouTube \\
& & Audiocaps-QA ~\citep{wang2025audiobench} & 0.3k & Multi-Domain\textsuperscript{$\ddagger$,$\diamond$} & AudioSet \\
& & MMAU ~\citep{sakshi2025mmau} & 10k & Multi-Domain\textsuperscript{$\ddagger$} & AudioSet and 12 others \\
\cmidrule(lr){2-6} 
& \multirow{1}{*}{\makecell[l]{LLM}} & 
EvalSIFT ~\citep{pandey2025sift} & 30k & Speech & Open-source ASR \\
\cmidrule(lr){2-6} 
& \multirow{1}{*}{\makecell[l]{MP-LLM\textsuperscript{$\dagger$}}} & 
MECAT-QA (Ours) & 20k & Extended~Multi-Domain\textsuperscript{$\S$} & ACAV100M \\
\bottomrule
\end{tabular}%
}
\vspace{-3mm}
\end{table*}

To ensure data source novelty, MECAT is constructed from a carefully selected subset of ACAV100M~\citep{lee2021acav100m}. 
This approach contrasts with benchmarks, such as AudioCaps~\citep{kim2019audiocaps}, Clotho~\citep{drossos2020clotho}, and WavCaps-QA~\citep{wang2025audiobench}, which predominantly draw from a limited pool of sources such as AudioSet~\citep{gemmeke2017audio} and Clotho~\citep{drossos2020clotho}~(see~\Cref{tab:benchmark_comparison}). 
The dataset comprises approximately 20,000 Creative Commons-licensed audio clips, each with a maximum duration of 10 seconds which is sufficient to contain one or a few salient acoustic events, while still allowing us to attach dense supervision for fine-grained, clip-local understanding.

Based on this unique data foundation, MECAT encompasses eight distinct audio domains designed to comprehensively represent real-world acoustic scenarios. These categories include four \textit{Pure} domains: silence (000), speech (S00),
\begin{figure}[!h]
  \includegraphics[width=\linewidth]{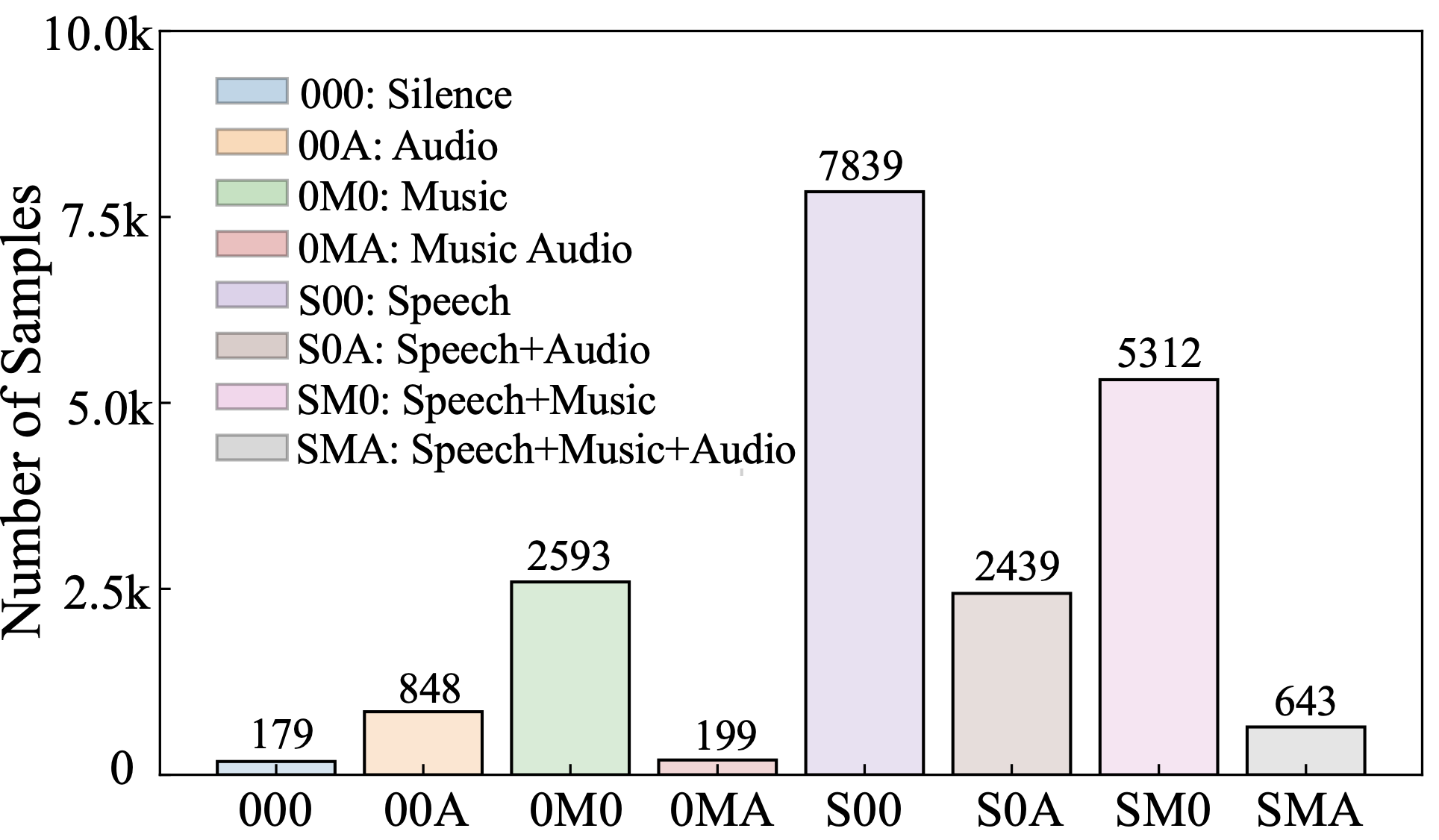}
  \caption{Distribution of audio samples across extended multi-domains in the MECAT, including speech, music, audio, combinations thereof, and silence.}
  \label{fig:audio_sample_distribution}
  \vspace{-3mm}
\end{figure} 
sound events (00A), and music (0M0), as well as all four possible combinations of \textit{Mixed} domains that reflect the complexity of natural auditory environments (SM0, S0A, 0MA, and SMA). 
This extended multi-domain coverage, with its distribution detailed in~\Cref{fig:audio_sample_distribution}, enables a nuanced evaluation of models on complex acoustic scenes, such as those that combine piano music with spoken discussion and audience applause.

\subsection{Tasks Definition}

As illustrated in~\Cref{fig:labeling-main}-B, the MECAT-Caption task delivers multi-perspective annotations for comprehensive evaluation. 
Each audio clip is annotated with a rich set of captions organized into three categories, which together comprise six distinct sub-categories. 
The first category, \textit{Systemic Captions}, consists of two sub-categories: a concise short caption focused on primary audio content and a detailed long caption encompassing contextual details and event interactions. 
The second category, \textit{Content-Specific Captions}, includes three sub-categories for the independent analysis of speech, music, and sound events. 
Crucially, to assess model performance across different levels of acoustic complexity, the evaluation for each content type is performed on corresponding pure domains (e.g., pure speech - S00) and all mixed domains.
Notably, these captions also explicitly state when a corresponding domain is absent. 
The final category is a single \textit{Content-Unrelated Caption} that focuses exclusively on acoustic characteristics like audio quality and reverberation. For each of these six sub-categories, three synonymous reference captions are provided, yielding a total of 18 reference captions per clip and creating a significantly richer vocabulary than existing datasets (see \Cref{appendix:B} for more details).

The final score $\text{Score}_{\text{Cap}}$ for the MECAT-Caption task is calculated as a weighted average of the three main categories:

\vspace{-3.5mm}
\begin{equation}
\begin{split}
\text{Score}_{\text{Cap}} = & \, 0.4 \cdot S_{\text{Systemic}} + 0.4 \cdot S_{\text{Content-Specific}} \\
& + 0.2 \cdot S_{\text{Content-Unrelated}}.
\end{split}
\vspace{-2mm}
\end{equation}

where the category scores are themselves weighted sums of their sub-categories:
\begin{align}
S_{\text{Systemic}} &= 0.8 \cdot S_{\text{Long}} + 0.2 \cdot S_{\text{Short}}, \\
S_{\text{Content-Specific}} &= 0.6 \cdot S_{\text{Speech}} + 0.3 \cdot S_{\text{Music}} + 0.1 \cdot S_{\text{Sound}}.
\vspace{-2mm}
\end{align}

The score for each content type ($S_{\text{Speech}}$, $S_{\text{Music}}$, $S_{\text{Sound}}$) is calculated as the unweighted mean of its performance on the corresponding pure domains (e.g., S00, 0M0, 00A) and all mixed domains. All coefficients are set heuristically to reflect the relative importance of overall scene, content detail, and acoustic context. Within the Content-Specific category, the $0.6/0.3/0.1$ weights roughly follow the relative prevalence of Speech, Music, and Sound in ACAV100M. A sensitivity analysis confirms that model rankings remain highly stable across alternative weighting scenarios (Kendall's $\tau = 0.92$; see \Cref{sec:weight_sensitivity}).

\begin{figure*}[ht]
  \centering
  \includegraphics[width=\linewidth]{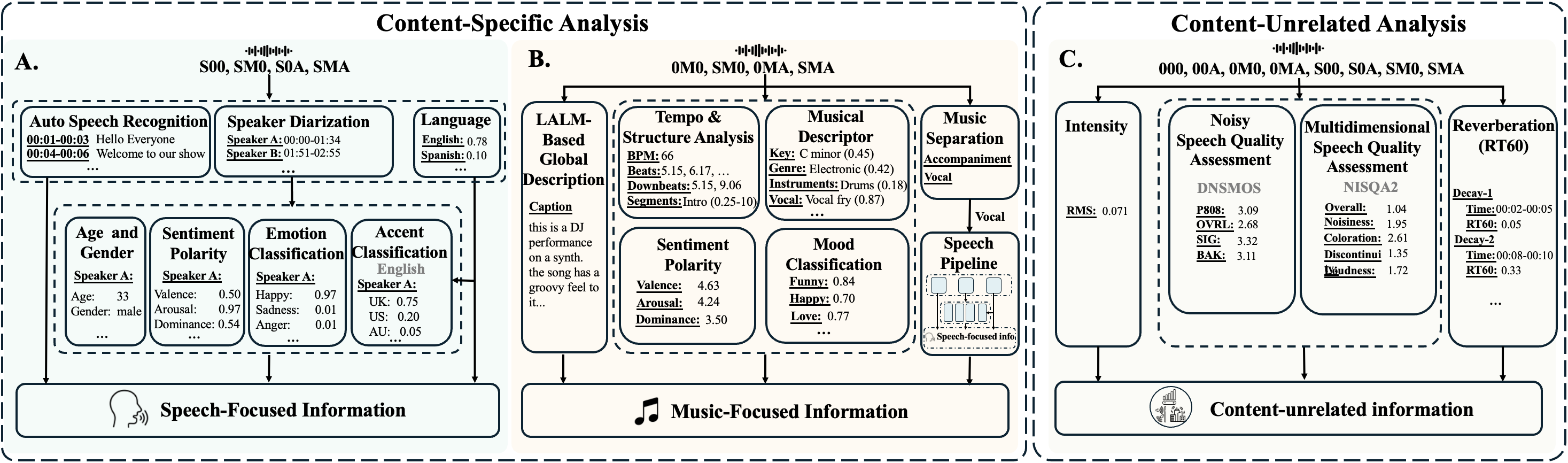}
  \caption{Domain Experts for Speech, Music, and Acoustic Properties.}
  \label{fig:annotation}
\end{figure*}

Complementing the captioning task, MECAT-QA facilitates evaluation through targeted, probing questions. Each audio clip is paired with five question-answer pairs that span different cognitive skills, resulting in over 100,000 QA pairs in total. These pairs are organized into three cognitive categories. The first, \textit{Perception}, focuses on the direct identification of audio content through its Direct Perception (DP) sub-category. 
The second, \textit{Analysis}, delves into acoustic properties via two sub-categories: Sound Characteristics (SC), for examining properties like pitch, and Quality Assessment (QAS), for evaluating technical fidelity. 
The final and most complex category, \textit{Reasoning}, targets higher-level cognitive skills through three sub-categories: Environment Reasoning (ER), requiring acoustic scene inference; Inference \& Judgement (IJ), involving logical deductions; and Application Context (AC), testing the understanding of practical scenarios.

The scoring for MECAT-QA is designed to ensure equal contribution from each cognitive skill. The overall score is the unweighted arithmetic mean of the scores from all six individual sub-categories:
\begin{equation}
\text{Score}_{\text{QA}} = (S_{\text{DP}} + S_{\text{SC}} + S_{\text{QAS}} + S_{\text{ER}} + S_{\text{IJ}} + S_{\text{AC}})/6.
\vspace{-2mm}
\end{equation}

\section{\vspace{-2pt}Annotation Construction}

This section details the MECAT annotation construction pipeline. 
As illustrated in \Cref{fig:labeling-main}, the process starts with a audio classification stage identifying the domain of each audio clip. 
Based on the resulting domains, the clip is then processed by a series of specialized expert models. 

The structured outputs from these experts are subsequently synthesized using LLM CoT reasoning to generate fine-grained captions and open-set QA pairs. 
The pipeline concludes with a rigorous quality control stage to ensure the reliability of all final annotations. 
The complete list of the used models is available in \Cref{appendix:C}.

\subsection{Domain Experts}
For each audio clip, we first use Audio Flamingo 2~\citep{ghosh2025audio} to generate a global, event-level summarization in natural language. Furthermore, we apply a series of domain expert models for more detailed analysis.

\noindent\textbf{Audio Classification}
\label{sec:pre-classification}
For each audio clip, we use CED-Base~\citep{dinkel2024ced} to predict AudioSet~\citep{gemmeke2017audio} labels for every 2-second, non-overlapping interval. 
This process results in a sequence of multi-label predictions for each clip. 
Based on the CED prediction, we categorize each clip into one of eight distinct domains: 000, 00A, 0M0, S00, SM0, 0MA, S0A, SMA, as detailed in Dataset Description Section.

\noindent\textbf{Speech-focused Analysis}
\label{sec:speech}
For speech-domain clips (S00, S0A, SM0, SMA), we employ a speech-focused analysis pipeline(\Cref{fig:annotation}-A). 
The pipeline consists of automatic speech recognition, language identification, and speaker diarization. 
Using the temporal boundaries from diarization, we extract each speaker's attributes, including gender, age, emotion, and English accent. The probabilities of these results are also utilized for subsequent LLM reasoning.

\noindent\textbf{Music-focused Analysis}
\label{sec:music}
For music-domain clips (0M0, SM0, 0MA, SMA), a music-focused analysis pipeline is employed (\Cref{fig:annotation}-B). It consists of LALM-based global description of music content (Audio Flamingo 2~\citep{ghosh2025audio}), musical attribute analysis, and music separation. Musical attribute analysis provides a series of perceptual and technical attributes such as emotions and tempo. The music separation module isolates vocal tracks from the instrumental background, which are then routed to the speech analysis pipeline.

\noindent\textbf{Sound Events-focused Analysis} For audios in 00A, we directly utilize the events labels predicted by the CED-Base model during the audio classification stage. 

\noindent\textbf{Acoustic Properties Analysis}
To extract fundamental signal characteristics and assess the recording environment, we apply a universal acoustic property analysis pipeline to all audio clips (Figure~\ref{fig:annotation}-C). The analysis content includes signal intensity, speech quality assessment, and reverberation.
Signal intensity is quantified via Root Mean Square (RMS). For audio quality, we conduct both DNSMOS~\citep{reddy2021dnsmos} and NISQA2~\citep{mittag2021nisqa} assessments to measure signal distortion, background noise, and perceptual quality. We also characterize the acoustic environment by estimating the reverberation time of the recording space.

\begin{figure*}[ht]
  \centering
  \includegraphics[width=0.80\textwidth]{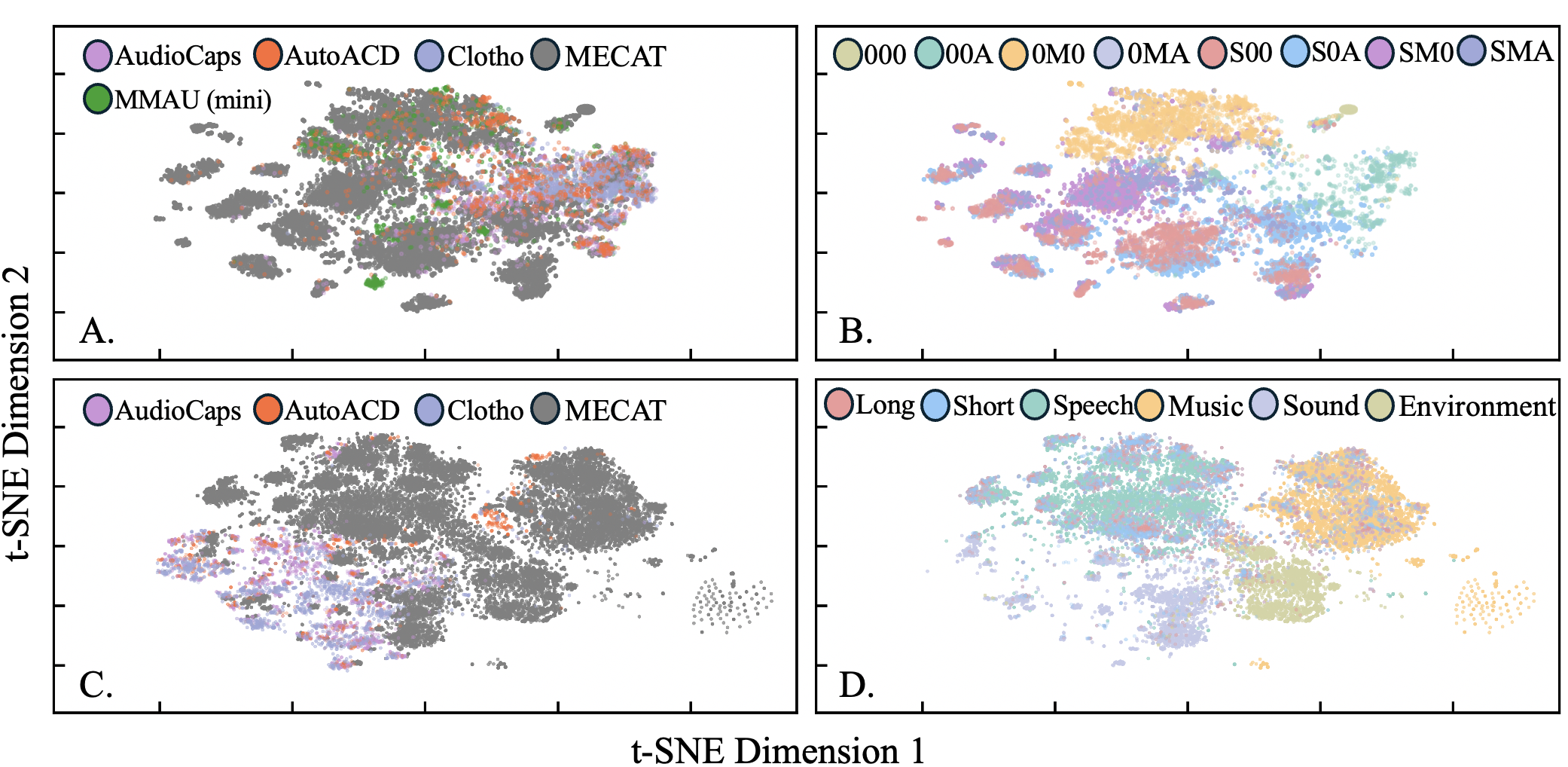}
  \caption{t-SNE plots of MECAT audio embeddings compared to other benchmarks (A), further clustered by domain (B). Caption embeddings are visualized in C and clustered by categories in D.
  Audio embeddings and captions embeddings are extracted from Dasheng-Base~\citep{dinkel2024dasheng} and Sentence-BERT~\citep{reimers2019sentence} respectively.}
  \label{fig:tsne}
  \vspace{-3mm}
\end{figure*}

\vspace{-7.15ex}
\subsection{LLM CoT Reasoning}
Our pipeline employs a Chain-of-Thought (CoT) guided LLM (Deepseek-R1: \citeauthor{guo2025deepseek},\citeyear{guo2025deepseek}) to synthesize a set of rich annotations. The model is instructed to reason over the outputs from all preceding analyses and the metadata. This reasoning process weighs evidence from various sources to resolve inconsistencies and identifying salient features. The final output consists of captions and corresponding QA pairs, where each item is annotated with a confidence level. The complete prompt is shown in \Cref{appendix:D}.
\subsection{Quality Control}
The model-based filtering use GLAP~\citep{dinkel2026glap} to compute the cosine similarity between audio clip and its systemic long caption embeddings. 
A sample is kept only if the similarity of its correct audio-caption pair exceeds its average similarity with a set of 6 other randomly selected captions by an empirically set threshold of 6.

We further apply rule-based filtering including LLM confidence thresholding, domain consistency between audio classification and LLM output, and hallucination removal~\citep{baranski2025investigation}.

\section{Metric Design}
\label{sec:date_metric}
Existing evaluation metrics
demonstrate significant limitations when evaluating fine-grained, detailed descriptions. 
To address this, we propose DATE, a metric built on Sentence-BERT~\citep{reimers2019sentence} that improves semantic assessment by combining single-sample semantic similarity and cross-sample discriminability score.

\noindent\textbf{Single-Sample Semantic Similarity} 
We apply embedding-level term frequency-inverse document frequency (TF-IDF) weighting to token embeddings from Sentence-BERT to emphasize tokens that are frequent within a single sample but rare across the dataset (details in \Cref{appendix:E}). The weighted embedding vector $\mathbf{v}_T$ for a given sentence $T$ is:  
\begin{equation}
\mathbf{v}_T = \sum_{t \in T} \big( \text{TF}_{\text{emb}}(t, T) \cdot \text{IDF}_{\text{emb}}(t) \big) \cdot E(t),
\vspace{-2mm}
\end{equation}
where $t$ is a token in $T$. The term $\text{TF}_{\text{emb}}(t, T)$, $\text{IDF}_{\text{emb}}(t)$, and $E(t)$ are the frequency, inverse document frequency, and the Sentence-BERT embedding, respectively.
The single-sample semantic similarity, $S_{{\rm sim}, i}$, is the cosine similarity between the weighted embeddings of the candidate and reference text:
\begin{equation}
    S_{\text{sim}, i} = (\mathbf{v}_{\text{cand}} \cdot \mathbf{v}_{\text{ref}}) / (\|\mathbf{v}_{\text{cand}}\| \|\mathbf{v}_{\text{ref}}\|).
  \vspace{-2mm}
\end{equation}

\begin{figure*}[ht]
  \centering
  \includegraphics[width=0.9\linewidth]{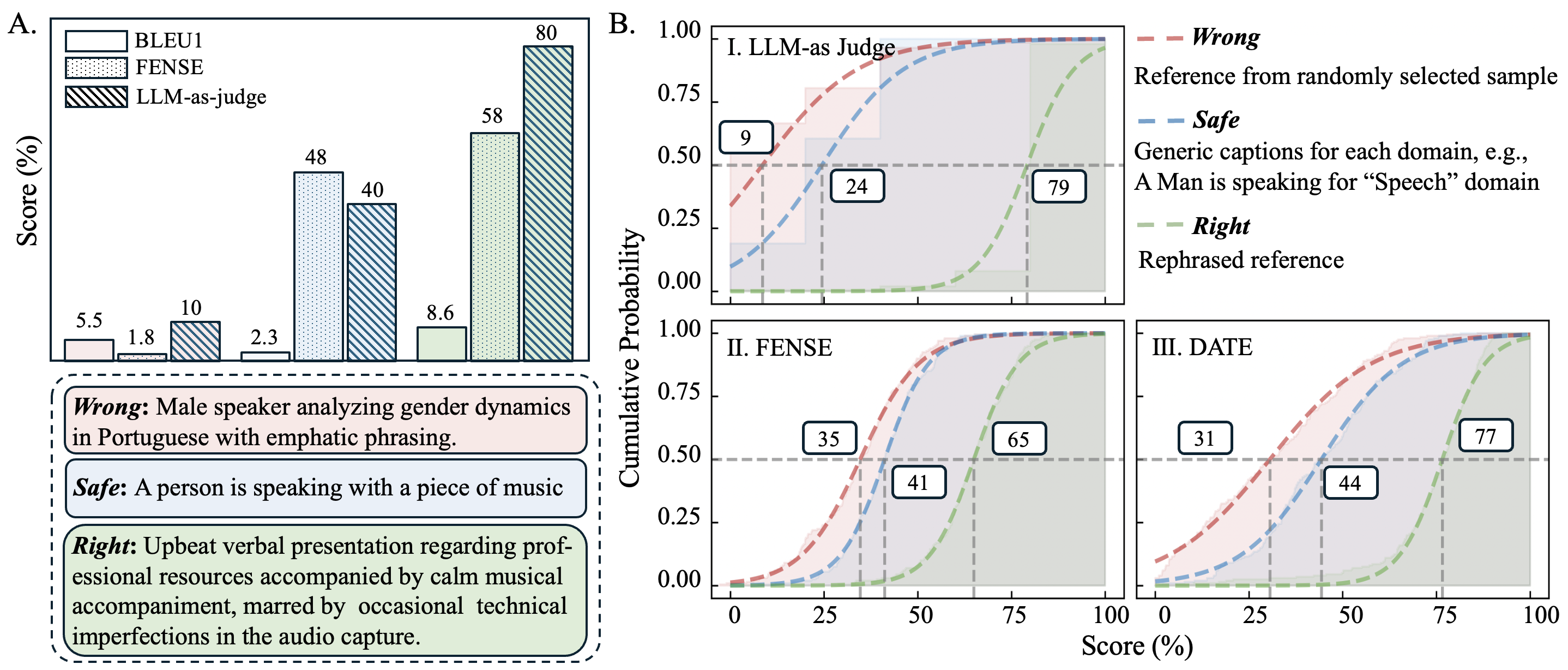}
  \caption{Metric Analysis. (A) Case study of existing metrics. Reference: ``An animated woman's voice shares information about learning materials while melodic instruments play quietly underneath, with persistent low-quality artifacts in the recording.'' (B) Cumulative Distribution Functions (CDF) of LLM-as-judge, FENSE, and DATE on Caption (left) and QA (right). Larger distances between CDF curves indicate better discriminative ability of the metric.}
  \label{fig:metric_analysis}
  \vspace{-3mm}
\end{figure*}

\noindent\textbf{Cross-Sample Discriminability} 
An ideal description should be clearly distinguishable from descriptions of other audio samples. 
We construct a cross-sample similarity matrix, $\mathcal{M}$, where each element $M_{i,j}$ is the score between the reference description for audio $i$ and the candidate description for audio $j$. 
For each sample $i$, we rank the correctly matched score $M_{i,i}$ against all candidate scores $\{M_{i,j}\}_{j=1}^N$. 
Denoting this rank as $r_i$, the discriminability score is:
\begin{equation}
S_{{\rm dis}, i} = 1 - r_i / N.
\vspace{-2mm}
\end{equation}
This rewards candidates that rank highly for their correct reference, approaching 1 for top ranks and 0 for bottom ranks.

\noindent\textbf{DATE} 
To ensures a balanced evaluation for both descriptive accuracy and uniqueness, the DATE score for each sample ${\rm DATE}_i$ is defined as the harmonic mean of its semantic similarity ($S_{\rm sim, i}$) and discriminability ($S_{\rm dis, i}$):
\begin{equation}
{\rm DATE}_i = \frac{2 \cdot S_{{\rm sim}, i} \cdot S_{{\rm dis}, i}}{S_{{\rm sim}, i} + S_{{\rm dis}, i}} \in [0, 1].
\vspace{-2mm}
\end{equation}
The DATE score of a dataset is $\frac{1}{N}\sum_{i=1}^N {\rm DATE}_i.$

\begin{table*}[h]
    \centering
    \caption{Model performance (DATE \%) on MECAT-Caption. \textbf{Bold} indicates the best performance, and \underline{underline} indicates the second best. $^\dagger$ indicates that its previous version (Audio Flamingo 2) was explicitly used in the data construction process.}
    \label{tab:bleu1_results}
    \resizebox{\textwidth}{!}{%
    \begin{tabular}{@{}llc*{3}{c}*{6}{c}c@{}}
    \toprule
    \multirow{4}{*}{Type} & 
    \multirow{4}{*}{Model} & 
    \multicolumn{2}{c}{Systemic} & 
    \multicolumn{6}{c}{Content-Specific} & 
    \multirow{3}{*}{\makecell[c]{Content\\Unrelated}} & 
    \multirow{4}{*}{Score$_\text{Cap}$} \\
    \cmidrule(lr){3-4}\cmidrule(lr){5-10} 
     & & 
    \multirow{2}{*}{\makecell[l]{Long}} & \multirow{2}{*}{\makecell[l]{Short}} & 
    \multicolumn{2}{c}{Speech} & 
    \multicolumn{2}{c}{Music} & 
    \multicolumn{2}{c}{Sound} & \\
    \cmidrule(lr){5-10} \cmidrule(lr){11-11} & & & & 
    Pure & Mixed & 
    Pure & Mixed & 
    Pure & Mixed & 
    \multicolumn{1}{c}{Env} & 
    \multicolumn{1}{c}{} & \\
    \midrule
    \multirow{2}{*}{\makecell[c]{Caption\\-Only}} 
     & Pengi & 43.5 & 46.8 & 27.2 & 29.5 & 29.3 & 13.1 & 42.8 & 14.6 & 7.1 & 29.4 \\
     & EnClap & 48.6 & 53.1 & 30.2 & 31.8 & 17.9 & 15.9 & 48.8 & 15.2 & 6.8 & 31.9 \\
    \midrule
    \multirow{15}{*}{LALM} 
     & Phi-4-Multimodal & 42.4 & 44.0 & 26.9 & 31.3 & 14.9 & 24.0 & 28.5 & 18.1 & 13.1 & 30.0 \\
     & Kimi-Audio-7B & 49.5 & 54.2 & 30.0 & 31.3 & 27.7 & 16.9 & 43.1 & 16.2 & 7.0 & 32.8 \\
     & Baichuan-Audio & 42.6 & 36.5 & 46.0 & 40.4 & 21.3 & 20.7 & 44.8 & 17.7 & 15.1 & 33.7 \\
     & Baichuan-Omni & 47.0 & 50.9 & 43.5 & 41.7 & 35.2 & 13.7 & 34.3 & 19.7 & 11.3 & 35.6 \\
     & MiMo-Audio & 56.5 & 56.9 & 45.8 & 44.9 & 35.8 & 19.4 & 46.8 & 21.0 & 9.8 & 40.1 \\
     & Audio Flamingo 3$^\dagger$ & 52.5 & 51.5 & 49.3 & 48.8 & 40.4 & 24.8 & 50.6 & 21.9 & 11.5 & 40.4 \\
     & Qwen3-Omni & 47.9 & 43.7 & 50.2 & 48.7 & 51.2 & 26.8 & 49.0 & 19.5 & 18.2 & 40.4 \\
     & Step-Audio-2-mini & 55.6 & 58.7 & 44.2 & 43.6 & 35.3 & 32.0 & 42.8 & 18.9 & 16.1 & 41.5 \\
     & Qwen2.5-Omni 3B & 56.4 & 55.2 & 42.5 & 41.3 & 46.6 & 29.7 & 52.9 & 23.9 & 19.4 & 42.5 \\
     & Qwen2.5-Omni 7B & 61.1 & 56.5 & 39.9 & 40.9 & 32.1 & 30.9 & 50.7 & 23.8 & 17.9 & 42.6 \\
     \cmidrule(lr){2-12}
     & Qwen3-Omni-Flash-1201 & \textbf{65.7} & 62.5 & 59.2 & 59.9 & \textbf{57.4} & 32.5 & \textbf{55.8} & \textbf{31.6} & \textbf{27.1} & 52.9 \\
     & Gemini-2.5-Flash & 65.6 & 63.9 & 57.5 & 57.5 & 52.9 & \textbf{41.0} & 52.2 & 28.3 & 22.1 & \underline{51.6} \\
     & Gemini-2.5-Pro & 62.3 & 62.4 & 56.6 & 57.5 & 53.6 & 38.7 & 53.4 & \underline{29.9} & 24.0 & 50.6 \\
     & Gemini-3-Flash & 63.6 & 61.9 & \underline{59.4} & \underline{60.8} & 43.1 & 32.9 & 51.1 & 29.7 & 25.7 & 51.1 \\
     & Gemini-3-Pro & \underline{64.9} & \textbf{65.8} & \textbf{60.5} & \textbf{62.4} & 49.8 & \underline{39.8} & \underline{55.1} & \underline{29.9} & \underline{26.1} & \textbf{53.1} \\
    \bottomrule
    \end{tabular}%
    }
    \vspace{-3mm}
    \end{table*}

\section{Result and Discussion}
This section presents an analysis of MECAT's data diversity, the analysis of the DATE metric, and a comprehensive evaluation of state-of-the-art models on MECAT.

\subsection{Data Analysis}

\noindent\textbf{Distribution and Diversity} The t-SNE analysis in \Cref{fig:tsne} highlights MECAT's superior coverage. Regarding \textit{audio} (\Cref{fig:tsne}-A/B), MECAT spans the full feature space with distinct internal clusters for pure domains, whereas existing benchmarks remain densely clustered around sound-event regions. Regarding \textit{captions} (\Cref{fig:tsne}-C/D), MECAT exhibits significantly richer semantic diversity driven by distinct content categories (speech, music, events) rather than simple length variations.

\noindent\textbf{Quality Validation} The trustworthiness of MECAT annotations is ensured through a rigorous multi-stage process. The expert models deployed in the pipeline (e.g., CED, Audio Flamingo 2) are architecturally independent with largely uncorrelated error distributions, making concurrent hallucination of the same content statistically unlikely. Combined with LLM-CoT conflict resolution and GLAP filtering, only ${\sim}$10\% of candidate samples survived the full pipeline for release. On this filtered set, two rounds of human evaluation were further conducted. A preference A/B test ($N=150$) verified that MECAT captions are superior to generic or inaccurate descriptions ($>94\%$ win rates) and comparable to human-written references (56.9\% win rate). Additionally, to directly assess dataset quality, 20 audio-domain professionals independently rated 700 samples (100 per category) on a 1--5 Relevance (REL) scale to evaluate the relevance between audio and provided captions, where 3 denotes a basic match and 1 denotes completely unrelated. The strict error rate (REL\,=\,1) was bounded at only 3.4\% [95\% CI: 2.3\%--5.1\%], and all seven categories scored significantly above 3.0 ($p < 0.001$, Wilcoxon test with Bonferroni correction). Comprehensive details are provided in \Cref{appendix:F}.

\begin{table*}[t]
    \centering
    \caption{Model Performance (DATE \%) on MECAT-QA. \textbf{Bold} indicates the best performance, and \underline{underline} indicates the second best. $^\dagger$ indicates that its previous version (Audio Flamingo 2) was explicitly used in the data construction process.}
    \label{tab:qa_results}
    \resizebox{\textwidth}{!}{%
    \begin{tabular}{@{}lc*{2}{c}*{3}{c}c@{}}
    \toprule
    \multirow{2}{*}{Model} & 
    \multicolumn{1}{c}{Perception} & 
    \multicolumn{2}{c}{Analysis} & 
    \multicolumn{3}{c}{Reasoning} & 
    \multirow{2}{*}{Score$_{\text{QA}}$} \\
    \cmidrule(lr){2-2}\cmidrule(lr){3-4}\cmidrule(lr){5-7}
     & 
    \makecell[c]{Direct\\Perception} & 
    \makecell[c]{Sound\\Characteristics} & 
    \makecell[c]{Quality\\Assessment} & 
    \makecell[c]{Environment\\Reasoning} & 
    \makecell[c]{Inference \&\\Judgment} & 
    \makecell[c]{Application\\Context} & 
     \\
    \midrule
     Kimi-Audio-7B & 45.6 & 39.2 & 18.7 & 34.6 & 48.9 & 41.2 & 38.0 \\
     Baichuan-Audio & 40.7 & 45.2 & 31.0 & 35.1 & 49.0 & 46.9 & 41.3 \\
     Baichuan-Omni & 43.6 & 44.7 & 33.7 & 39.9 & 49.3 & 49.1 & 43.4 \\
     Phi-4-Multimodal & 48.4 & 46.3 & 34.7 & 40.2 & 49.3 & 48.7 & 44.6 \\
     MiMo-Audio & \underline{59.3} & 49.3 & 24.9 & 39.1 & 52.7 & 46.2 & 45.2 \\
     Step-Audio-2-mini & 57.7 & 54.3 & 37.2 & 39.2 & 48.9 & 48.0 & 47.6 \\
     Audio Flamingo 3$^\dagger$ & 53.8 & 50.2 & 36.0 & 43.0 & 54.5 & 49.6 & 47.8 \\
     Qwen2.5-Omni 3B & 55.7 & 53.2 & 38.6 & 41.1 & 51.8 & 50.8 & 48.5 \\
     Qwen2.5-Omni 7B & 57.8 & 52.9 & \underline{39.1} & 44.0 & 53.2 & 50.8 & 49.6 \\
     Qwen3-Omni & \textbf{61.7} & \underline{54.6} & \textbf{39.3} & 45.0 & 56.9 & 56.1 & \textbf{52.3} \\
     \cmidrule(lr){1-8}
     Qwen3-Omni-Flash-1201 & 48.0 & 45.9 & 29.5 & 45.6 & 56.7 & 54.8 & 46.7 \\
     Gemini-2.5-Flash & 56.3 & \textbf{55.3} & 37.7 & 46.8 & \textbf{58.6} & \textbf{58.0} & \underline{52.1} \\ 
     Gemini-2.5-Pro & 55.5 & 54.4 & 37.7 & \textbf{47.6} & \underline{57.3} & 56.6 & 51.5 \\
     Gemini-3-Flash & 54.3 & 51.1 & 34.1 & \underline{47.2} & 57.2 & \underline{57.0} & 51.0 \\
     Gemini-3-Pro & 55.5 & 45.5 & 25.8 & 44.0 & 53.2 & 52.0 & 46.0 \\
    \bottomrule
    \end{tabular}%
    }
    \vspace{-3mm}
    \end{table*}

\subsection{Metric Analysis}
\label{sec:metric_analysis}

This section validates our proposed metric, DATE, against the strong baseline FENSE, using an LLM-as-judge method as the upper-bound reference (prompts and reliability analysis in \Cref{appendix:G,appendix:H}).

\noindent\textbf{Qualitative Analysis} The case study in \Cref{fig:metric_analysis}-A exposes critical flaws in existing metrics. Lexical-based metrics like BLEU-1 are semantically unreliable, assigning higher scores to Wrong captions than Safe ones. While FENSE improves upon this, it struggles to distinguish high-quality (Right) from vague (Safe) captions, showing a negligible score gap ($\Delta \approx 10$). In contrast, DATE aligns with the clear separation observed in LLM-as-judge scores.
 DATE demonstrates a clear advantage over FENSE, evidenced by significantly larger median score spans for both Right vs. Wrong (DATE: 46 vs. FENSE: 30) and Right vs. Safe (DATE: 33 vs. FENSE: 24).

\noindent\textbf{Quantitative Analysis} The Cumulative Distribution Function (CDF) curves in \Cref{fig:metric_analysis}-B further quantify discriminative power, where larger inter-curve distances indicate superior performance. Furthermore, a bootstrap stability analysis confirms that DATE produces reliable scores, with Coefficient of Variation (CV) below 1\% at full scale for nearly all evaluation groups, and CV dropping below 3\% at 5\% sampling. More information could be seen in \Cref{appendix:F}.

\noindent\textbf{Alignment with Human Judgment} To assess the alignment between DATE and human perception, we utilized the same 150 A/B caption pairs described in \Cref{appendix:F}. Captions preferred by human evaluators received substantially higher DATE scores (Mean: 90.9) compared to non-preferred ones (Mean: 49.3). This significant margin indicates that DATE is strongly correlated with human preferences regarding accuracy and detail, as shown in \Cref{tab:date_human_alignment}.

\subsection{Model Performance on MECAT}
\subsubsection{Overall Performance}

An extensive collection of publicly available models was evaluated, including 17 models for Captioning (15 LALMs, 2 traditional baselines) and 15 LALMs for QA. The evaluated models are strictly categorized into two primary types: traditional Caption-Only models (e.g., Pengi~\citep{NEURIPS2023_3a2e5889}, EnClap~\citep{10446672}) and Large Audio-Language Models (LALMs). The LALMs are further stratified into four architectural subcategories: i) \textit{Audio-focused LALMs} (e.g., Kimi-Audio~\citep{ding2025kimi}, MiMo-Audio~\citep{zhang2025mimo}, Baichuan-Audio~\citep{li2025baichuan}, Step-Audio-2-mini~\citep{wu2025stepaudio2}, Audio Flamingo 3~\citep{ghosh2026audio}), ii) \textit{Omni LALMs} (e.g., Qwen2.5-Omni~\citep{xu2025qwen2}, Qwen3-Omni~\citep{xu2025qwen3}, Qwen3-Omni-Flash-1201~\citep{qwen2025flash1201}, Baichuan-Omni~\citep{li2024baichuan}), iii) \textit{Multimodal LALMs} (Phi-4-Multimodal~\citep{phi42025}), and iv) the \textit{Gemini series}~\citep{comanici2025gemini,google2025gemini3}. Detailed specifications regarding the model architectures and the corresponding prompts utilized in this study are provided in \Cref{appendix:I}.

\noindent\textbf{Performance on MECAT-Caption.} As shown in \Cref{tab:bleu1_results}, LALMs demonstrate a substantial advantage over traditional baselines, with scores ranging from 29.4 (Pengi) to 53.1 (Gemini-3-Pro), attributed to superior instruction-following capabilities. 
Regarding domain stability, a distinct disparity is observed: while performance on Speech tasks remains robust, Music and Sound tasks suffer significant degradation ($10\%\sim25\%$) when transitioning from Pure to Mixed setting. This suggests a prevalent speech-centric bias in current architectures. Furthermore, the consistently suboptimal performance on Content-Unrelated tasks underscores an urgent need for models to better capture intrinsic sound properties beyond high-level event recognition.

\noindent\textbf{Performance on MECAT-QA.} The hierarchy in the QA task (\Cref{tab:qa_results}) largely mirrors captioning, yet with a notable shift: the gap between proprietary and open-weight models narrows significantly. Qwen3-Omni (52.3) slightly outperforms both Gemini-2.5-Flash (52.1) and Pro (51.5). At a granular level, we observe a distinct \textit{capability dichotomy}: models excel in Direct Perception and content-based Inference, but degrade significantly on tasks requiring the analysis of intrinsic acoustic properties, such as Quality Assessment and Environment Reasoning. For instance, even top models fail to exceed a score of 40 in Quality Assessment, confirming that current LALMs prioritize high-level semantics over nuanced acoustic interpretation.

\subsubsection{In-depth Analysis: LALM Bottlenecks}

While standard metrics rank models, they often mask underlying behavioral nuances. This study utilizes the Captioning task to probe two critical bottlenecks: discriminability limitations and robustness against hallucination.

\noindent\textbf{The Critical Role of Discriminability.} A granular analysis of the Pure Speech subset using DATE (\Cref{appendix:J}) underscores the interplay between Semantic Similarity and Discriminability (note that for other subtasks, discriminability could be derived from the Similarity and DATE scores in \Cref{tab:bleu1_results} and \Cref{appendix:K}). Top-tier proprietary models---notably Qwen3-Omni-Flash-1201 and the Gemini series---dominate both metrics, pairing superior similarity with exceptional discriminability (above 77.0). Notably, the open-source Qwen3-Omni demonstrates that robust discriminative power (64.7) can compensate for moderate similarity to secure a top-tier ranking. This suggests that while elite models excel globally, strong acoustic discriminability is a primary driver for high-quality captioning.


\noindent\textbf{Hallucination in Silent Segments.} Qualitative evaluation of silent segments (\Cref{appendix:L}) reveals a significant robustness issue. While models like Qwen3-Omni-Flash-1201 and Gemini-2.5  correctly identify silence, many LALMs over-generate, resulting in hallucinations of specific but unrelated text (e.g., ``\textit{I'm gonna be a daddy}'' or ``\textit{Thank you}''). This tendency to over-generate exposes a fundamental vulnerability, underscoring the necessity for improved rejection mechanisms in future architectures.

\subsection{Limitations}
While MECAT advances fine-grained audio evaluation, several limitations should be acknowledged.

\textbf{Audio scope.} The 10-second clip duration prioritizes dense, local-level supervision over long-range temporal reasoning; MECAT is thus complementary to long-audio benchmarks such as LongAudioBench rather than a substitute.

\textbf{Annotation quality.} Although the multi-expert pipeline and rigorous filtering substantially mitigate upstream model errors, residual noise from expert hallucinations cannot be entirely eliminated, particularly in non-speech categories where annotation difficulty is higher.

\textbf{DATE metric.} DATE inherits fundamental limitations of embedding-based similarity, including potential insensitivity to entity swaps, and its cross-sample discriminability component is most informative when the test set contains diverse audio clips; on highly homogeneous corpora, its discriminative power may diminish.

\section{Conclusion and Future Work}
In this work, we introduced MECAT, a Multi-Experts Constructed Benchmark leveraged by Chain-of-Thought reasoning to advance fine-grained audio understanding in Captioning and QA tasks. Complementing this, we proposed DATE, a novel metric tailored to penalize vague terminology and incentivize detailed, discriminative descriptions.

\clearpage
\section*{Acknowledgements}
We specially thank Jiahao Mei (Shanghai Jiao Tong University), who contributed to the subjective evaluation during the revision process. His work on experimental design and data collection greatly improved the quality of this paper. We also thank all annotators who participated in the evaluation, and the anonymous reviewers for their constructive feedback.

\section*{Impact Statement}
This work contributes a benchmark and metric for audio understanding evaluation. We discuss the potential societal implications below.

\textbf{Positive impact.} MECAT and DATE are designed to advance the evaluation of audio understanding systems, which can yield positive societal benefits such as improved accessibility for hearing-impaired individuals and enhanced context-aware environmental awareness.

\textbf{Potential risks.} Fine-grained audio captioning systems could potentially be applied to automatically describe audio recordings containing private or sensitive information, raising concerns about privacy leakage and surveillance misuse.

\textbf{Mitigation.} MECAT itself is constructed exclusively from publicly available, Creative Commons-licensed data (ACAV100M), ensuring the benchmark does not contain private content. For downstream deployment of fine-grained audio captioning systems, we recommend that practitioners implement privacy-preserving safeguards, including sensitive content detection and filtering before caption generation, anonymization of personally identifiable information in outputs, and adherence to informed consent principles when processing non-public audio data.


\bibliography{example_paper}

@inproceedings{papineni2002bleu,
  title={Bleu: a method for automatic evaluation of machine translation},
  author={Papineni, Kishore and Roukos, Salim and Ward, Todd and Zhu, Wei-Jing},
  booktitle={Proceedings of the 40th Annual Meeting of the Association for Computational Linguistics},
  pages={311--318},
  year={2002}
}

@inproceedings{vedantam2015cider,
  title={Cider: Consensus-based image description evaluation},
  author={Vedantam, Ramakrishna and Lawrence Zitnick, C and Parikh, Devi},
  booktitle={Proceedings of the IEEE/CVF Conference on Computer Vision and Pattern Recognition},
  pages={4566--4575},
  year={2015}
}

@inproceedings{anderson2016spice,
  title={Spice: Semantic propositional image caption evaluation},
  author={Anderson, Peter and Fernando, Basura and Johnson, Mark and Gould, Stephen},
  booktitle={Proceedings of the European Conference on Computer Vision (ECCV)},
  pages={382--398},
  year={2016},
  organization={Springer}
}

@inproceedings{gemmeke2017audio,
  title={Audio Set: An Ontology and Human-labeled Dataset for Audio Events},
  author={Gemmeke, Jort F and Ellis, Daniel PW and Freedman, Dylan and Jansen, Aren and Lawrence, Wade and Moore, R Channing and Plakal, Manoj and Ritter, Marvin},
  booktitle={Proceedings of the IEEE International Conference on Acoustics, Speech and Signal Processing (ICASSP)},
  pages={776--780},
  year={2017},
  organization={IEEE}
}

@book{lyon2017human,
  title={Human and machine hearing},
  author={Lyon, Richard F},
  year={2017},
  publisher={Cambridge University Press}
}

@inproceedings{kim2019audiocaps,
  title={Audiocaps: Generating captions for audios in the wild},
  author={Kim, Chris Dongjoo and Kim, Byeongchang and Lee, Hyunmin and Kim, Gunhee},
  booktitle={Proceedings of the 2019 Conference of the North American Chapter of the Association for Computational Linguistics: Human Language Technologies, Volume 1 (Long and Short Papers)},
  pages={119--132},
  year={2019}
}

@inproceedings{wu2019audiocaption_listen_and_tell,
  title={Audio caption: Listen and tell},
  author={Wu, Mengyue and Dinkel, Heinrich and Yu, Kai},
  booktitle={Proceedings of the IEEE International Conference on Acoustics, Speech and Signal Processing (ICASSP)},
  pages={830--834},
  year={2019},
  organization={IEEE}
}

@inproceedings{drossos2020clotho,
  title={Clotho: an Audio Captioning Dataset},
  author={Drossos, Konstantinos and Lipping, Samuel and Virtanen, Tuomas},
  booktitle={Proceedings of the IEEE International Conference on Acoustics, Speech and Signal Processing (ICASSP)},
  pages={736--740},
  year={2020},
  organization={IEEE}
}

@inproceedings{kong2021decoupling,
  title={Decoupling Magnitude and Phase Estimation with Deep ResUNet for Music Source Separation.},
  author={Kong, Qiuqiang and Cao, Yin and Liu, Haohe and Choi, Keunwoo and Wang, Yuxuan },
  booktitle={Proceedings of the 22nd International Society for Music Information Retrieval Conference (ISMIR)},
  year={2021},
  pages={342--349},
  organization={Citeseer}
}

@inproceedings{lee2021acav100m,
  title={Acav100m: Automatic curation of large-scale datasets for audio-visual video representation learning},
  author={Lee, Sangho and Chung, Jiwan and Yu, Youngjae and Kim, Gunhee and Breuel, Thomas and Chechik, Gal and Song, Yale},
  booktitle={Proceedings of the IEEE/CVF International Conference on Computer Vision},
  pages={10274--10284},
  year={2021}
}

@inproceedings{mittag2021nisqa,
  title={NISQA: A Deep CNN-Self-Attention Model for Multidimensional Speech Quality Prediction with Crowdsourced Datasets},
  author={Mittag, Gabriel and Naderi, Babak and Chehadi, Assmaa and M{\"o}ller, Sebastian},
  booktitle={Proceedings of the 22nd Interspeech Conference (Interspeech)},
  pages={2127--2131},
  year={2021}
}

@inproceedings{reddy2021dnsmos,
  title={DNSMOS: A non-intrusive perceptual objective speech quality metric to evaluate noise suppressors},
  author={Reddy, Chandan KA et al.},
  booktitle={Proceedings of the IEEE International Conference on Acoustics, Speech and Signal Processing (ICASSP)},
  pages={6493--6497},
  year={2021},
  organization={IEEE}
}

@article{speechbrain,
  author  = {Mirco Ravanelli and Titouan Parcollet and Adel Moumen and Sylvain de Langen and Cem Subakan and Peter Plantinga and Yingzhi Wang and Pooneh Mousavi and Luca Della Libera and Artem Ploujnikov and Francesco Paissan and Davide Borra and Salah Zaiem and Zeyu Zhao and Shucong Zhang and Georgios Karakasidis and Sung-Lin Yeh and Pierre Champion and Aku Rouhe and Rudolf Braun and Florian Mai and Juan Zuluaga-Gomez and Seyed Mahed Mousavi and Andreas Nautsch and Ha Nguyen and Xuechen Liu and Sangeet Sagar and Jarod Duret and Salima Mdhaffar and Ga{{\"e}}lle Laperri{{\`e}}re and Mickael Rouvier and Renato De Mori and Yannick Est{{\`e}}ve},
  title   = {Open-Source Conversational AI with SpeechBrain 1.0},
  journal = {Journal of Machine Learning Research},
  year    = {2024},
  volume  = {25},
  number  = {333},
  pages   = {1--11},
}

@inproceedings{li2022learning,
  title={Learning to answer questions in dynamic audio-visual scenarios},
  author={Li, Guangyao and Wei, Yake and Tian, Yapeng and Xu, Chenliang and Wen, Ji-Rong and Hu, Di},
  booktitle={Proceedings of the IEEE/CVF Conference on Computer Vision and Pattern Recognition},
  pages={19108--19118},
  year={2022}
}

@inproceedings{lipping2022clotho,
  title={Clotho-aqa: A crowdsourced dataset for audio question answering},
  author={Lipping, Samuel and Sudarsanam, Parthasaarathy and Drossos, Konstantinos and Virtanen, Tuomas},
  booktitle={Proceedings of the 30th European Signal Processing Conference (EUSIPCO)},
  pages={1140--1144},
  year={2022},
  organization={IEEE}
}

@inproceedings{reddy2022dnsmos,
  title={DNSMOS P. 835: A non-intrusive perceptual objective speech quality metric to evaluate noise suppressors},
  author={Reddy, Chandan KA et al.},
  booktitle={Proceedings of the IEEE International Conference on Acoustics, Speech and Signal Processing (ICASSP)},
  pages={886--890},
  year={2022},
  organization={IEEE}
}

@inproceedings{zhou2022fense,
  title={Can audio captions be evaluated with image caption metrics?},
  author={Zhou, Zelin and Zhang, Zhiling and Xu, Xuenan and Xie, Zeyu and Wu, Mengyue and Zhu, Kenny Q},
  booktitle={Proceedings of the IEEE International Conference on Acoustics, Speech and Signal Processing (ICASSP)},
  pages={981--985},
  year={2022},
  organization={IEEE}
}

@inproceedings{baranski2025investigation,
  title={Investigation of whisper asr hallucinations induced by non-speech audio},
  author={Bara{\'n}ski, Mateusz and Jasi{\'n}ski, Jan and Bartolewska, Julitta and Kacprzak, Stanis{\l}aw and Witkowski, Marcin and Kowalczyk, Konrad},
  booktitle={Proceedings of the IEEE International Conference on Acoustics, Speech and Signal Processing (ICASSP)},
  pages={1--5},
  year={2025},
  organization={IEEE}
}

@inproceedings{burkhardt2023speech,
  title={Speech-based age and gender prediction with transformers},
  author={Burkhardt, Felix and Wagner, Johannes and Wierstorf, Hagen and Eyben, Florian and Schuller, Bj{\"o}rn},
  booktitle={Speech Communication; 15th ITG Conference},
  pages={46--50},
  year={2023},
  organization={VDE}
}

@inproceedings{bredin2023pyannote,
  title={Pyannote. audio 2.1 speaker diarization pipeline: principle, benchmark, and recipe},
  author={Bredin, Herv{\'e}},
  booktitle={Proceedings of the 24th Interspeech Conference (Interspeech)},
  pages={1983--1987},
  year={2023},
  organization={ISCA}
}

@inproceedings{doh2023lp,
  title={{LP-MusicCaps}: LLM-Based Pseudo Music Captioning},
  author={Doh, Seungheon and Nam, Juhan},
  booktitle={Proceedings of the 24th International Society for Music Information Retrieval Conference},
  year={2023},
  organization={International Society for Music Information Retrieval Conference}
}

@inproceedings{ghosh2025audio,
  title={{Audio Flamingo 2}: An Audio-Language Model with Long-Audio Understanding and Expert Reasoning Abilities},
  author={Sreyan Ghosh and Zhifeng Kong and Sonal Kumar and S Sakshi and Jaehyeon Kim and Wei Ping and Rafael Valle and Dinesh Manocha and Bryan Catanzaro},
  booktitle={Proceedings of the 40th International Conference on Machine Learning (ICML)},
  year={2025},
  pages={1--48},
}

@misc{hawley2023shaart,
  title={{SHAART}: Speech and Hearing in Audio and Real Time},
  author={Hawley, Scott H.},
  year={2023},
  howpublished={\url{https://github.com/drscotthawley/SHAART}},
}

@inproceedings{li2023mert,
  title={{MERT}: Acoustic music understanding model with large-scale self-supervised training},
  author={Yizhi, LI and Yuan, Ruibin and Zhang, Ge and Ma, Yinghao and Chen, Xingran and Yin, Hanzhi and Xiao, Chenghao and Lin, Chenghua and Ragni, Anton and Benetos, Emmanouil and others},
  booktitle={Proceedings of the International Conference on Learning Representations (ICLR)},
  pages={1--24},
  year={2023}
}

@inproceedings{radford2023whisper,
  title={Robust speech recognition via large-scale weak supervision},
  author={Radford, Alec and Kim, Jong Wook and Xu, Tao and Brockman, Greg and McLeavey, Christine and Sutskever, Ilya},
  booktitle={Proceedings of the 40th International Conference on Machine Learning (ICML)},
  pages={28492--28518},
  year={2023}
}

@inproceedings{sakshi2025mmau,
  title={{MMAU}: A Massive Multi-Task Audio Understanding and Reasoning Benchmark},
  author={Sakshi, S and Tyagi, Utkarsh and Kumar, Sonal and Seth, Ashish and Selvakumar, Ramaneswaran and Nieto, Oriol and Duraiswami, Ramani and Ghosh, Sreyan and Manocha, Dinesh},
  booktitle={Proceedings of the International Conference on Learning Representations (ICLR)},
  pages = {1--36},
  year={2025}
}

@inproceedings{taejun2023allinone,
  title={All-In-One Metrical And Functional Structure Analysis With Neighborhood Attentions on Demixed Audio},
  author={Kim, Taejun and Nam, Juhan},
  booktitle={Proceedings of the IEEE Workshop on Applications of Signal Processing to Audio and Acoustics (WASPAA)},
  year={2023}
}

@article{wagner2023dawn,
  title={Dawn of the transformer era in speech emotion recognition: closing the valence gap},
  author={Wagner, Johannes and Triantafyllopoulos, Andreas and Wierstorf, Hagen and Schmitt, Maximilian and Burkhardt, Felix and Eyben, Florian and Schuller, Bj{\"o}rn W},
  journal={IEEE Transactions on Pattern Analysis and Machine Intelligence},
  volume={45},
  number={9},
  pages={10745--10759},
  year={2023},
  publisher={IEEE}
}

@inproceedings{zuluaga2023commonaccent,
  title={CommonAccent: Exploring Large Acoustic Pretrained Models for Accent Classification Based on Common Voice},
  author={Zuluaga-Gomez, Juan and Ahmed, Sara and Visockas, Danielius and Subakan, Cem},
  booktitle={Proceedings of the 24th Interspeech Conference (Interspeech)},
  pages={5291--5295},
  year={2023},
  organization={ISCA}
}

@article{zheng2023judging,
  title={Judging {LLM}-as-a-judge with {MT-bench} and {Chatbot Arena}},
  author={Zheng, Lianmin and Chiang, Wei-Lin and Sheng, Ying and Zhuang, Siyuan and Wu, Zhanghao and Zhuang, Yonghao and Lin, Zi and Li, Zhuohan and Li, Dacheng and Xing, Eric and others},
  journal={Advances in Neural Information Processing Systems},
  volume={36},
  pages={46595--46623},
  year={2023}
}

@inproceedings{dinkel2024ced,
  title={{CED}: Consistent ensemble distillation for audio tagging},
  author={Dinkel, Heinrich and Wang, Yongqing and Yan, Zhiyong and Zhang, Junbo and Wang, Yujun},
  booktitle={Proceedings of the IEEE International Conference on Acoustics, Speech and Signal Processing (ICASSP)},
  pages={291--295},
  year={2024},
  organization={IEEE}
}

@inproceedings{hu2024wavllm,
  title={{WavLLM}: Towards Robust and Adaptive Speech Large Language Model},
  author={Hu, Shujie and Zhou, Long and Liu, Shujie and Chen, Sanyuan and Meng, Lingwei and Hao, Hongkun and Pan, Jing and Liu, Xunying and Li, Jinyu and Sivasankaran, Sunit and others},
  booktitle={Proceedings of the Findings of the Association for Computational Linguistics (EMNLP)},
  pages={4552--4572},
  year={2024}
}

@inproceedings{dinkel2024dasheng,
  title={Scaling up masked audio encoder learning for general audio classification},
  author={Dinkel, Heinrich and Yan, Zhiyong and Wang, Yongqing and Zhang, Junbo and Wang, Yujun and Wang, Bin},
  booktitle={Proceedings of the 25th Interspeech Conference (Interspeech)},
  pages={547--551},
  year={2024}
}

@inproceedings{huang2024audiogpt,
  title={{AudioGPT}: Understanding and generating speech, music, sound, and talking head},
  author={Huang, Rongjie and Li, Mingze and Yang, Dongchao and Shi, Jiatong and Chang, Xuankai and Ye, Zhenhui and Wu, Yuning and Hong, Zhiqing and Huang, Jiawei and Liu, Jinglin and others},
  booktitle={Proceedings of the AAAI Conference on Artificial Intelligence},
  volume={38},
  pages={23802--23804},
  year={2024}
}

@inproceedings{lee2024fleur,
  title={{FLEUR}: An Explainable Reference-Free Evaluation Metric for Image Captioning Using a Large Multimodal Model},
  author={Lee, Yebin and Park, Imseong and Kang, Myungjoo},
  booktitle={Proceedings of the 62nd Annual Meeting of the Association for Computational Linguistics (Volume 1: Long Papers)},
  pages={3732--3746},
  year={2024}
}

@inproceedings{liu2024enhancing,
  title={Enhancing Automated Audio Captioning via Large Language Models with Optimized Audio Encoding},
  author={Liu, Jizhong and Li, Gang and Zhang, Junbo and Dinkel, Heinrich and Wang, Yongqing and Yan, Zhiyong and Wang, Yujun and Wang, Bin},
  booktitle={Proceedings of the 25th Interspeech Conference (Interspeech)},
  pages={1135--1139},
  year={2024}
}

@article{liu2024leveraging,
  title={Leveraging {CED} encoder and large language models for automated audio captioning},
  author={Liu, Jizhong and Li, Gang and Zhang, Junbo and Liu, Chenyu and Dinkel, Heinrich and Wang, Yongqing and Yan, Zhiyong and Wang, Yujun and Wang, Bin},
  journal={Proceedings of the DCASE Challenge},
  pages={1--4},
  year={2024}
}

@article{mei2024wavcaps,
  title={{WavCaps}: A chatgpt-assisted weakly-labelled audio captioning dataset for audio-language multimodal research},
  author={Mei, Xinhao and Meng, Chutong and Liu, Haohe and Kong, Qiuqiang and Ko, Tom and Zhao, Chengqi and Plumbley, Mark D and Zou, Yuexian and Wang, Wenwu},
  journal={IEEE/ACM Transactions on Audio, Speech, and Language Processing},
  volume={32},
  pages={3339--3354},
  year={2024},
  publisher={IEEE}
}

@inproceedings{sun2024auto,
  title={{Auto-ACD}: A large-scale dataset for audio-language representation learning},
  author={Sun, Luoyi and Xu, Xuenan and Wu, Mengyue and Xie, Weidi},
  booktitle={Proceedings of the 32nd ACM International Conference on Multimedia},
  pages={5025--5034},
  year={2024}
}

@inproceedings{tang2023salmonn,
  title={{SALMONN}: Towards Generic Hearing Abilities for Large Language Models},
  author={Tang, Changli and Yu, Wenyi and Sun, Guangzhi and Chen, Xianzhao and Tan, Tian and Li, Wei and Lu, Lu and MA, Zejun and Zhang, Chao},
  booktitle={Proceedings of the International Conference on Learning Representations (ICLR)},
  pages={1--23},
  year={2024}
}

@inproceedings{wang2025audiobench,
  title={{AudioBench}: A Universal Benchmark for Audio Large Language Models},
  author={Wang, Bin and Zou, Xunlong and Lin, Geyu and Sun, Shuo and Liu, Zhuohan and Zhang, Wenyu and Liu, Zhengyuan and Aw, Aiti and Chen, Nancy},
  booktitle={Proceedings of the 2025 Conference of the Nations of the Americas Chapter of the Association for Computational Linguistics: Human Language Technologies (Volume 1: Long Papers)},
  pages={4297--4316},
  year={2025}
}

@inproceedings{yuan2025sound,
  title={{Sound-VECaps}: Improving Audio Generation with Visually Enhanced Captions},
  author={Yuan, Yi and Jia, Dongya and Zhuang, Xiaobin and Chen, Yuanzhe and Chen, Zhuo and Wang, Yuping and Wang, Yuxuan and Liu, Xubo and Kang, Xiyuan and Plumbley, Mark D and others},
  booktitle={Proceedings of the IEEE International Conference on Acoustics, Speech and Signal Processing (ICASSP)},
  pages={1--5},
  year={2025},
  organization={IEEE}
}

@book{plack2023sense,
  title={The sense of hearing},
  author={Plack, Christopher J},
  year={2023},
  publisher={Routledge}
}

@article{chu2023qwen,
  title={{Qwen-Audio}: Advancing Universal Audio Understanding via Unified Large-Scale Audio-Language Models},
  author={Chu, Yunfei and Xu, Jin and Zhou, Xiaohuan and Yang, Qian and Zhang, Shiliang and Yan, Zhijie and Zhou, Chang and Zhou, Jingren},
  journal={arXiv preprint arXiv:2311.07919},
  year={2023},
}

@article{du2023lauragpt,
  title={{LauraGPT}: Listen, Attend, Understand, and Regenerate Audio with {GPT}},
  author={Du, Zhihao and Wang, Jiaming and Chen, Qian and Chu, Yunfei and Gao, Zhifu and Li, Zerui and Hu, Kai and Zhou, Xiaohuan and Xu, Jin and Ma, Ziyang and others},
  journal={arXiv preprint arXiv:2310.04673},
  year={2023},
}

@inproceedings{manco2023song,
  title={{The Song Describer Dataset}: a Corpus of Audio Captions for Music-and-Language Evaluation},
  author={Manco, Ilaria and Weck, Benno and Doh, Seungheon and Won, Minz and Zhang, Yixiao and Bogdanov, Dmitry and Wu, Yusong and Chen, Ke and Tovstogan, Philip and Benetos, Emmanouil and others},
  booktitle={NeurIPS Machine Learning for Audio Workshop},
  year={2023}
}

@article{rubenstein2023audiopalm,
  title={{AudioPaLM}: A Large Language Model That Can Speak and Listen},
  author={Rubenstein, Paul K. and Asawaroengchai, Chulayuth and Nguyen, Duc Dung and Bapna, Ankur and Borsos, Zal{\'a}n and Quitry, F{\'e}lix de Chaumont and Chen, Peter and Badawy, Dalia El and Han, Wei and Kharitonov, Eugene and others},
  journal={arXiv preprint arXiv:2306.12925},
  year={2023},
}

@article{shu2023llasm,
  title={{LLASM}: Large Language and Speech Model},
  author={Shu, Yu and Dong, Siwei and Chen, Guangyao and Huang, Wenhao and Zhang, Ruihua and Shi, Daochen and Xiang, Qiqi and Shi, Yemin},
  journal={arXiv preprint arXiv:2308.15930},
  year={2023},
}

@article{wang2023blsp,
  title={{BLSP}: Bootstrapping Language-Speech Pre-Training via Behavior Alignment of Continuation Writing},
  author={Wang, Chen and Liao, Minpeng and Huang, Zhongqiang and Lu, Jinliang and Wu, Junhong and Liu, Yuchen and Zong, Chengqing and Zhang, Jiajun},
  journal={arXiv preprint arXiv:2309.00916},
  year={2023},
}

@article{chen2023x,
  title={{X-LLM}: Bootstrapping Advanced Large Language Models by Treating Multi-Modalities as Foreign Languages},
  author={Chen, Feilong and Han, Minglun and Zhao, Haozhi and Zhang, Qingyang and Shi, Jing and Xu, Shuang and Xu, Bo},
  journal={arXiv preprint arXiv:2305.04160},
  year={2023},
}

@inproceedings{ma2024emotion2vec,
  title={emotion2vec: Self-supervised pre-training for speech emotion representation},
  author={Ma, Ziyang and Zheng, Zhisheng and Ye, Jiaxin and Li, Jinchao and Gao, Zhifu and Zhang, Shiliang and Chen, Xie},
  booktitle={Findings of the Association for Computational Linguistics: ACL 2024},
  pages={15747--15760},
  year={2024}
}

@article{kang2025towards,
  title={Towards Unified Music Emotion Recognition across Dimensional and Categorical Models},
  author={Kang, Jaeyong and Herremans, Dorien},
  journal={arXiv preprint arXiv:2502.03979},
  year={2025}
}

@inproceedings{ma2025mmar,
 author = {Ma, Ziyang and Ma, Yinghao and Zhu, Yanqiao and Yang, Chen and Chao, Yi-Wen and Xu, Ruiyang and Chen, Wenxi and Chen, Yuanzhe and Chen, Zhuo and Cong, Jian and Li, Kai and Li, Keliang and Li, Siyou and Li, Xinfeng and Li, Xiquan and Lian, Zheng and Liang, Yuzhe and Liu, Minghao and Niu, Zhikang and Wang, Tianrui and Yuping, Wang and Wang, Yuxuan and Wu, Yihao and Yang, Guanrou and Yu, Jianwei and Yuan, Ruibin and Zheng, Zhisheng and Zhou, Ziya and Zhu, Haina and Xue, Wei and Benetos, Emmanouil and Yu, Kai and Chng, Eng-Siong and Chen, Xie},
 booktitle = {Advances in Neural Information Processing Systems},
 editor = {D. Belgrave and C. Zhang and H. Lin and R. Pascanu and P. Koniusz and M. Ghassemi and N. Chen},
 pages = {},
 publisher = {Curran Associates, Inc.},
 title = {{MMAR}: A Challenging Benchmark for Deep Reasoning in Speech, Audio, Music, and Their Mix},
 volume = {38},
 year = {2025}
}

@article{pandey2025sift,
  title={{SIFT-50M}: A large-scale multilingual dataset for speech instruction fine-tuning},
  author={Pandey, Prabhat and Swaminathan, Rupak Vignesh and Girish, KV and Sen, Arunasish and Xie, Jian and Strimel, Grant P and Schwarz, Andreas},
  journal={arXiv preprint arXiv:2504.09081},
  year={2025}
}

@article{guo2025deepseek,
  title={{DeepSeek-R1} incentivizes reasoning in LLMs through reinforcement learning},
  author={Guo, Daya and Yang, Dejian and Zhang, Haowei and Song, Junxiao and Wang, Peiyi and Zhu, Qihao and Xu, Runxin and Zhang, Ruoyu and Ma, Shirong and Bi, Xiao and others},
  journal={Nature},
  volume={645},
  number={8081},
  pages={633--638},
  year={2025},
  publisher={Nature Publishing Group UK London}
}

@inproceedings{dinkel2026glap,
  title={{GLAP}: General contrastive audio-text pretraining across domains and languages},
  author={Dinkel, Heinrich and Yan, Zhiyong and Wang, Tianzi and Wang, Yongqing and Sun, Xingwei and Niu, Yadong and Liu, Jizhong and Li, Gang and Zhang, Junbo and Luan, Jian},
  booktitle={Proceedings of the IEEE International Conference on Acoustics, Speech and Signal Processing (ICASSP)},
  pages={14737--14741},
  year={2026},
  organization={IEEE}
}

@inproceedings{reimers2019sentence,
  title={{Sentence-BERT}: Sentence Embeddings using {Siamese BERT-Networks}},
  author={Reimers, Nils and Gurevych, Iryna},
  booktitle={Proceedings of the 2019 Conference on Empirical Methods in Natural Language Processing and the 9th International Joint Conference on Natural Language Processing (EMNLP-IJCNLP)},
  pages={3982--3992},
  year={2019}
}

@article{xu2025qwen2,
  title={{Qwen2.5-Omni} technical report},
  author={Xu, Jin and Guo, Zhifang and He, Jinzheng and Hu, Hangrui and He, Ting and Bai, Shuai and Chen, Keqin and Wang, Jialin and Fan, Yang and Dang, Kai and others},
  journal={arXiv preprint arXiv:2503.20215},
  year={2025}
}

@article{ding2025kimi,
  title={{Kimi-Audio} technical report},
  author={Ding, Ding and Ju, Zeqian and Leng, Yichong and Liu, Songxiang and Liu, Tong and Shang, Zeyu and Shen, Kai and Song, Wei and Tan, Xu and Tang, Heyi and others},
  journal={arXiv preprint arXiv:2504.18425},
  year={2025}
}

@INPROCEEDINGS{10446672,
  author={Kim, Jaeyeon and Jung, Jaeyoon and Lee, Jinjoo and Woo, Sang Hoon},
  booktitle={Proceedings of the IEEE International Conference on Acoustics, Speech and Signal Processing (ICASSP)}, 
  title={{EnCLAP}: Combining Neural Audio Codec and Audio-Text Joint Embedding for Automated Audio Captioning}, 
  year={2024},
  volume={},
  number={},
  pages={6735-6739},
  doi={10.1109/ICASSP48485.2024.10446672}}

@inproceedings{NEURIPS2023_3a2e5889,
 author = {Deshmukh, Soham and Elizalde, Benjamin and Singh, Rita and Wang, Huaming},
 booktitle = {Advances in Neural Information Processing Systems},
 editor = {A. Oh and T. Naumann and A. Globerson and K. Saenko and M. Hardt and S. Levine},
 pages = {18090--18108},
 publisher = {Curran Associates, Inc.},
 title = {Pengi: An Audio Language Model for Audio Tasks},
 volume = {36},
 year = {2023}
}

@article{phi42025,
  title={Phi-4-mini technical report: Compact yet powerful multimodal language models via mixture-of-loras},
  author={Abouelenin, Abdelrahman and Ashfaq, Atabak and Atkinson, Adam and Awadalla, Hany and Bach, Nguyen and Bao, Jianmin and Benhaim, Alon and Cai, Martin and Chaudhary, Vishrav and Chen, Congcong and others},
  journal={arXiv preprint arXiv:2503.01743},
  year={2025}
}

@article{li2025baichuan,
  title={{Baichuan-Audio}: A unified framework for end-to-end speech interaction},
  author={Li, Tianpeng and Liu, Jun and Zhang, Tao and Fang, Yuanbo and Pan, Da and Wang, Mingrui and Liang, Zheng and Li, Zehuan and Lin, Mingan and Dong, Guosheng and others},
  journal={arXiv preprint arXiv:2502.17239},
  year={2025}
}

@article{li2024baichuan,
  title={{Baichuan-Omni} technical report},
  author={Li, Yadong and Sun, Haoze and Lin, Mingan and Li, Tianpeng and Dong, Guosheng and Zhang, Tao and Ding, Bowen and Song, Wei and Cheng, Zhenglin and Huo, Yuqi and others},
  journal={arXiv preprint arXiv:2410.08565},
  year={2024}
}

@article{zhang2025mimo,
  title={{MiMo-Audio}: Audio Language Models are Few-Shot Learners},
  author={Zhang, Dong and Wang, Gang and Xue, Jinlong and Fang, Kai and Zhao, Liang and Ma, Rui and Ren, Shuhuai and Liu, Shuo and Guo, Tao and Zhuang, Weiji and others},
  journal={arXiv preprint arXiv:2512.23808},
  year={2025}
}

@article{ghosh2026audio,
  title={{Audio Flamingo 3}: Advancing audio intelligence with fully open large audio language models},
  author={Ghosh, Sreyan and Goel, Arushi and Kim, Jaehyeon and Kumar, Sonal and Kong, Zhifeng and Lee, Sang-gil and Yang, Chao-Han and Duraiswami, Ramani and Manocha, Dinesh and Valle, Rafael and others},
  journal={Advances in Neural Information Processing Systems},
  volume={38},
  pages={41819--41886},
  year={2026}
}

@article{xu2025qwen3,
  title={{Qwen3-Omni} technical report},
  author={Xu, Jin and Guo, Zhifang and Hu, Hangrui and Chu, Yunfei and Wang, Xiong and He, Jinzheng and Wang, Yuxuan and Shi, Xian and He, Ting and Zhu, Xinfa and others},
  journal={arXiv preprint arXiv:2509.17765},
  year={2025}
}

@article{comanici2025gemini,
  title={{Gemini 2.5}: Pushing the frontier with advanced reasoning, multimodality, long context, and next generation agentic capabilities},
  author={Comanici, Gheorghe and Bieber, Eric and Schaekermann, Mike and Pasupat, Ice and Sachdeva, Noveen and Dhillon, Inderjit and Blistein, Marcel and Ram, Ori and Zhang, Dan and Rosen, Evan and others},
  journal={arXiv preprint arXiv:2507.06261},
  year={2025}
}

@article{wu2025stepaudio2,
  title={{Step-Audio 2} technical report},
  author={Wu, Boyong and Yan, Chao and Hu, Chen and Yi, Cheng and Feng, Chengli and Tian, Fei and Shen, Feiyu and Yu, Gang and Zhang, Haoyang and Li, Jingbei and others},
  journal={arXiv preprint arXiv:2507.16632},
  year={2025}
}

@misc{google2025gemini3,
  title={A new era of intelligence with {Gemini} 3},
  author={Google},
  year={2025},
  howpublished={\url{https://blog.google/products-and-platform/products/gemini/gemini-3/}}
}

@misc{qwen2025flash1201,
  title={{Hear You. See You. Follow Smarter!}},
  author={{Qwen Team}},
  year={2025},
  howpublished={\url{https://qwen.ai/blog?id=qwen3-omni-flash-20251201}}
}
\bibliographystyle{icml2026}

\newpage
\appendix
\crefalias{section}{appendix}
\numberwithin{equation}{section}
\numberwithin{table}{section}
\numberwithin{figure}{section}
\onecolumn
\section{LLM usage}
\label{appendix:A}
In the preparation of this manuscript, we utilized Gemini 2.5 Pro model (accessed in July 2025) primarily for proofreading and grammatical corrections. The tool was used to improve the clarity and readability of the text. All authors have reviewed and edited the final manuscript and take full responsibility for its content.

\section{Vocabulary Size of Audio Captioning TestSet}
\label{appendix:B}
This section introduces vocabulary size comparison to demonstrate the lexical diversity of MECAT-Caption. The following table indicates that the vocabulary size of MECAT-Caption contains about 4-17 times more words than the existing dataset. 

\begin{table}[hb]
    \centering
    \caption{Vocabulary Size of Audio Captioning Test Set}
    \label{tab:vocabulary_comparison}
    \begin{tabular}{rl}
        \toprule
        {Dataset}       & {\# Vocab} \\ 
        \midrule
        AudioCaps              & 5,581             \\
        AudioCaps-Enhanced     & 1,260             \\
        AutoACD                & 3,517             \\
        Clotho                 & 1,852             \\
        StrongDescriber        & 2,726             \\
        LPMusicCaps-MTT        & 1,666             \\
        \midrule
        \textbf{MECAT-Caption} & \textbf{22,595}   \\
        \bottomrule
    \end{tabular}
\end{table}

\clearpage
\section{Deployed Acoustic Models in Processing Pipeline}
\label{appendix:C}

This section introduces the acoustic models deployed in our processing pipeline. These models are categorized into Content-Specific models (including Speech, Music, and Sound analysis) and Content-Unrelated models (Environment analysis), each designed to handle different aspects of audio understanding tasks.

\begin{table*}[h]
\centering
\caption{Deployed Acoustic Models in Processing Pipeline}
\resizebox{\linewidth}{!}{%
\label{tab:pipeline_models}
\begin{tabular}{llll}
\toprule
Category & Subcategory & Model           & Analsysis Task                                                                    \\ \midrule
\multirow{13}{*}{\makecell[l]{Content\\Specific}}& \multirow{7}{*}{Speech} & Speechbrain-ECAPA\citep{speechbrain}       & Language Recognition                                                                                  \\
                                   &                         & Whisper Large v2 \citep{radford2023whisper}        & Auto Speech Recognition                                                                    \\
                                   &                         & Pyannote-SD 3.1 \citep{bredin2023pyannote}          & Speaker Diariazation                                                                       \\
                                   &                         & Emotion2Vec \citep{ma2024emotion2vec}           & Speaker Emotion Recognition                                                                \\
                                   &                         & Audeering-DSER \citep{wagner2023dawn}        & Dimentional Speaker Emotion Recogintion                                                    \\
                                   &                         & Audeering-AGR \citep{burkhardt2023speech}         & Age and Gender Recoginition                                                                \\
                                   &                         & CommonAccent \citep{zuluaga2023commonaccent}            & English Accent Recognition                                                                 \\ \cmidrule(lr){2-4}
                                   & \multirow{5}{*}{Music}  & Music Structure Analyzer \citep{taejun2023allinone} & Tempo \& Structure                                                                         \\
                                   &                         & Music2Emo \citep{kang2025towards}               & Emotion (Sentiment Polarity and Mood) \\
                                   &                         & MERT \citep{li2023mert}                   & Musical Descriptor                                                                         \\
                                   &                         & ByteSep \citep{kong2021decoupling}                 & Music Seperation                                                                           \\
                                   &                         & Audio Flamingo 2 \citep{ghosh2025audio}         & AudioLLM                                                                                   \\ \cmidrule(lr){2-4}
                                   & Sound                   & CED \citep{dinkel2024ced}                      & Sound Event Recognition                                                                    \\ \midrule
\multirow{3}{*}{\makecell[l]{Content\\Unrelated}} &  \multirow{3}{*}{\makecell[l]{Environment}}       & DNSMOS \citep{reddy2021dnsmos,reddy2022dnsmos}                  & Noisy Speech Quality Assessment                                                            \\
                                   &                         & NISQA V2.0 \citep{mittag2021nisqa}              & Multidimensional Speech Quality Assessment                                                 \\
                                   &                         & SHAART \citep{hawley2023shaart}                  & Reverberation                                                                              \\ \bottomrule
\end{tabular}
}
\end{table*}

\clearpage
\section{LLM Audio Analysis Synthesis Prompts}
\label{appendix:D}
\subsection*{\textnormal{Act as an expert audio analysis synthesizer to process multi-model JSON outputs through this workflow}}

\subsection*{\textnormal{Step 1: Multi-Domain Data Specifications}}

\subsubsection*{\textnormal{1.1 Multi-Domain Input Integration}}

\begin{enumerate}[label=\alph*), leftmargin=3em]
\item Speech: Speech recognition, speech emotion, speaker diarization and so on
\item Music: Structure analysis, technical descriptors, emotion and so on
\item Sound: Event detection timestamps, classifications
\item Environment: Acoustic characteristics, interference markers
\item Meta-info: Title and description of original video where audio clip was extracted
\end{enumerate}

\subsubsection*{\textnormal{1.2 Data Integrity Challenges}}
\begin{enumerate}[label=\alph*), leftmargin=3em]
\item Missing fields
\item Contradictory model outputs
\item Confidence score variances
\end{enumerate}

\subsection*{\textnormal{Step 2: Technical and Analytical Limitations}}

\subsubsection*{\textnormal{2.1 Model and System Constraints}}
\begin{enumerate}[label=\alph*), leftmargin=3em]
\item No speech recognition ability in audio captioning models (e.g., audio-flamingo variants)
\item Accuracy disparities across the analyzed domains
\item Potential conflicting information between models
\end{enumerate}

\subsubsection*{\textnormal{2.2 Audio Content Heterogeneity}}
\begin{enumerate}[label=\alph*), leftmargin=3em]
\item Hybrid audio types (e.g., speech, music, sound-event, environment)
\item Variable audio properties (e.g., clip lengths or quality)
\item Reliable topic or domain, but absent or non-relevant details in Meta-info
\end{enumerate}

\subsection*{\textnormal{Step 3: Audio Analysis Workflow}}

\subsection*{\textnormal{3.1 Salient Feature Identification}}

\quad 3.1.1 Identify dominant characteristics of this audio:

~~What makes this specific audio clip unique according to the analysis? Examples include:
\begin{enumerate}[label=\alph*), leftmargin=3em]
\item Specific spoken phrases
\item Dominant musical styles or moods
\item Significant sound events
\item The overall acoustic scene
\item Notable quality issues
\item Complex interplay of elements
\end{enumerate}

3.1.2 Supporting Evidence Extraction:

~~Gather the key details describing these salient features from the relevant JSON fields

\subsection*{\textnormal{3.2 Synthesis Rules}}

\quad 3.2.1 Generation Rules:

\begin{enumerate}[label=\alph*), leftmargin=3em]
\item Critically weigh evidence from different fields, considering inaccuracies or conflicts and accounting for domain-specific limitations
\item Prioritize information most reliable or central to the audio's character based on overall data patterns
\item Carefully identify conflicting information between fields and avoid mentioning conflicting aspects in the final caption. Focus only on consistent and unopposed information. Do not invent details not present in the data
\item Crucial Constraint:
\begin{itemize}[leftmargin=1em]
\item The final generated text must strictly describe only the analyzed content of the audio segment itself
\item It must not refer to the topic, title, description, or inferred subject matter from the overall video metadata
\item Avoid phrases like "in a clip from a video about..." or similar references to the source video's topic
\item Prohibit using parentheses to provide detailed explanation in any output, e.g., Moderate tempo (88 BPM)
\end{itemize}
\end{enumerate}

\quad 3.2.2 Perspective Rules:

\quad ~~ALL answers must be created from the perspective of someone who ONLY LISTENED to the 

\quad ~~audio without any technical/model references or quantitative metrics (e.g., BPM, MOS, etc.)

\quad 3.2.3 Evaluation Rules:

\quad ~~Assign a confidence level (High or Low) based on the following aspects:
\begin{enumerate}[label=\alph*), leftmargin=3em]
\item Consistency: Are the different analyses in the JSON generally consistent or contradictory? High consistency increases confidence
\item Completeness: Is key information present? (Fewer gaps = higher confidence)
\item Clarity: How clearly does the consistent data point to the audio's nature? (Less ambiguity in reliable data = higher confidence)
\item Metadata Context Usefulness: How relevant and useful was the overall video metadata in confirming or contextualizing findings from the clip's direct analysis?
\end{enumerate}

\subsection*{\textnormal{3.3 Caption Development Framework}}

\quad 3.3.1 Systematic Caption

\begin{enumerate}[label=\alph*), leftmargin=3em]
\item Short ($<$ 15 words):
\begin{itemize}[leftmargin=1em]
\item Protocol: Primary domain characteristics + Most prevalent characteristic from cross-model correlation
\item Example: Blues guitar performance at live concert with audience reactions
\end{itemize}
\item Long (1-2 sentences):
\begin{itemize}[leftmargin=1em]
\item Protocol: Primary domain + significant secondary elements + notable quality factors
\item Example: A live concert recording featuring guitar with crowd cheers, despite occasional microphone static
\end{itemize}
\end{enumerate}

\quad 3.3.2 Content-Focused Caption

\begin{enumerate}[label=\alph*), leftmargin=3em]
\item Speech:
\begin{itemize}[leftmargin=1em]
\item Protocol: ASR content + paralinguistic context
\item Example: Two speakers discussing jazz history, with piano accompaniment
\end{itemize}
\item Music:
\begin{itemize}[leftmargin=1em]
\item Protocol: Technical descriptors + performance context
\item Example: Upbeat electronic track with distant traffic noise
\end{itemize}
\item Sound:
\begin{itemize}[leftmargin=1em]
\item Protocol: Event taxonomy + spatial relationships
\item Example: Office environment with printer hum and keyboard typing, mild echo present
\end{itemize}
\end{enumerate}

\quad 3.3.3 Content-Unrelated Caption

\begin{enumerate}[label=\alph*), leftmargin=3em]
\item Environment:
\begin{itemize}[leftmargin=1em]
\item Protocol: Acoustic properties + interference profile
\item Example: Studio recording with noticeable background interference
\end{itemize}
\end{enumerate}

\quad 3.3.4 Caption Variants

\begin{enumerate}[label=\alph*), leftmargin=3em]
\item Lexical substitution (WordNet-based synonyms)
\item Structural reordering (active/passive voice)
\item Descriptive equivalence ('crowd cheers' $\rightarrow$ 'audience applause')
\end{enumerate}

\quad 3.3.5 Null Handling

\quad ~~When no domain-specific elements are detected:
\begin{enumerate}[label=\alph*), leftmargin=3em]
\item Use explicit 'None' declaration in content field
\item Generate null statement variants (e.g., 'No discernible speech content', 'Musical elements appear absent')
\end{enumerate}

\subsection*{3.4 Question-Answering Design}

\quad 3.4.1 Content Categories

\quad ~~Include questions across:
\begin{enumerate}[label=\alph*), leftmargin=3em]
\item Direct Perception (sound type, volume, duration)
\item Sound Characteristics (timbre, rhythm, frequency characteristics)
\item Environmental Perception (recording setting, echo, background noise)
\item Quality Assessment (clarity, interference factors)
\item Inference and Judgment (sound source, generation method, object properties)
\item Application Context (use cases, semantic meaning)
\end{enumerate}

\quad 3.4.2 Difficulty Levels

\quad ~~Include a mix of:
\begin{enumerate}[label=\alph*), leftmargin=3em]
\item Basic: Direct descriptive questions (e.g., 'What sound is heard?')
\item Intermediate: Analytical questions (e.g., 'What are the characteristics of this sound?')
\item Advanced: Inferential questions (e.g., 'In what environment was this recorded?')
\item Complex: Comprehensive judgment questions (e.g., 'Based on the sound, what is the most likely material?')
\end{enumerate}

\quad 3.4.3 Question Distribution

\quad ~~Basic (25\%) | Intermediate (35\%) | Advanced (25\%) | Complex (15\%)

\quad 3.4.4 Question Variety

\quad ~~Include:
\begin{enumerate}[label=\alph*), leftmargin=3em]
\item Ensure questions cover all listed categories
\item Avoid repetitive question patterns or formats
\item Include both yes/no questions and open-ended questions
\item Include some questions about what is NOT present in the audio
\item Include some comparative questions (e.g., 'Does this sound more like X or Y?')
\end{enumerate}

\quad 3.4.5 Cognitive Levels

\quad ~~Include:
\begin{enumerate}[label=\alph*), leftmargin=3em]
\item Include questions requiring simple recognition
\item Include questions requiring analysis of components
\item Include questions requiring synthesis of information
\item Include questions requiring evaluation or judgment
\end{enumerate}

\subsection*{\textnormal{Step 4: Structured Output Specification (JSON Format)}}

\quad ~~Confidence: High/Low

\quad ~~Possible Conflicts: None or list of conflicting fields

\quad ~~Reasoning: 2-3 line evaluation considering model consensus and data quality

\quad ~~Short-Caption: Single-sentence essence

\quad ~~Short-Caption-Variants-1: Paraphrased version 1

\quad ~~Short-Caption-Variants-2: Paraphrased version 2

\quad ~~Main-Caption: Integrated summary

\quad ~~Main-Caption-Variants-1: Paraphrased version 1

\quad ~~Main-Caption-Variants-2: Paraphrased version 2

\quad ~~Speech-Captions: Speech-focused analysis or NONE

\quad ~~Speech-Caption-Variants-1: Paraphrased version 1

\quad ~~Speech-Caption-Variants-2: Paraphrased version 2

\quad ~~Music-Captions: Music-focused analysis or NONE

\quad ~~Music-Caption-Variants-1: Paraphrased version 1

\quad ~~Music-Caption-Variants-2: Paraphrased version 2

\quad ~~Sound-Captions: Sound-focused analysis or NONE

\quad ~~Sound-Caption-Variants-1: Paraphrased version 1

\quad ~~Sound-Caption-Variants-2: Paraphrased version 2

\quad ~~Environment-Caption: Environment-focused analysis

\quad ~~Environment-Caption-Variants-1: Paraphrased version 1

\quad ~~Environment-Caption-Variants-2: Paraphrased version 2

\quad ~~QA-Pair-1-id: 1 or None

\quad ~~QA-Pair-1-difficulty: basic, intermediate, advanced, or complex

\quad ~~QA-Pair-1-category: direct perception,sound characteristics, environmental perception,quality assessment, inference judgment,or application context

\quad ~~QA-Pair-1-question: question content

\quad ~~QA-Pair-1-answer: answer content

\quad ~~QA-Pair-2-id: 2 or None

\quad ~~QA-Pair-2-difficulty: basic, intermediate, advanced, or complex

\quad ~~QA-Pair-2-category: direct perception,sound characteristics, environmental perception,quality assessment, inference judgment,or application context

\quad ~~QA-Pair-2-question: question content 

\quad ~~QA-Pair-2-answer: answer content

\quad ~~// ... 3 more QA pairs following the same pattern

\clearpage
\section{Embedding-level TF-IDF Calculation for DATE}
\label{appendix:E}

\subsection{Rationale for Embedding-level TF-IDF}

Classific term frequency-inverse document frequency (TF-IDF) relies on discrete, hard-count token occurrences to calculate term frequency (TF). However, in natural language, the semantic relationship between words (e.g., "dog" and "canine") is lost when treating them as independent tokens.

The Discriminability based Audio Task Evaluation (DATE) metric enhances the representational power of sentence embeddings by incorporating an Embedding-level TF-IDF weighting scheme. This method leverages the semantic information encoded in word embeddings to compute a soft, non-integer Term Frequency, thereby improving the quality of the resulting sentence representation for downstream similarity and discrimination calculations.

\subsection{Method of Calculation}

The Embedding-level TF-IDF weight for a word $w$ in a sentence $s$ is calculated as the product of its semantic-aware Term Frequency ($\text{TF}_{\text{emb}}$) and its Inverse Document Frequency ($\text{IDF}_{\text{emb}}$):

\begin{equation}
\text{TF-IDF}_{\text{emb}}(w, s) = \text{TF}_{\text{emb}}(w, s) \times \text{IDF}_{\text{emb}}(w)
\end{equation}

The calculation proceeds in three main steps:

\noindent\textbf{Semantic-aware Term Frequency ($\text{TF}_{\text{emb}}$)}
Instead of counting a word's exact occurrences (which would result in an integer count), $\text{TF}_{\text{emb}}$ is calculated by measuring the average similarity of the word's embedding to the embeddings of all other words within the same sentence. This accounts for semantic context and relatedness.

\begin{itemize}
    \item \textbf{Word Embeddings:} Word-level embeddings are generated for all tokens in the corpus with the Sentense-Bert.
    \item \textbf{Word-to-Word Similarity Matrix ($\text{WordSim}$):} A full $\text{WordSim}$ matrix is computed by taking the dot product of the $L_2$-normalized embeddings ($E$) of all unique non-padding words in the batch.
    \begin{equation}
    \text{WordSim} = \text{Normalize}(E) \cdot \text{Normalize}(E)^T
    \end{equation}
    \item \textbf{Term Frequency Calculation:} The semantic Term Frequency for a word $w_i$ in a sentence $s$ is calculated by summing the squared similarity scores with all other words $w_j$ in that sentence (excluding special tokens like \texttt{[CLS]} and \texttt{[SEP]}):
    \begin{equation}
    \text{TF}_{\text{emb}}(w_i, s) = \sum_{w_j \in s, j \neq i} \text{WordSim}(i, j)^2
    \end{equation}
    \item \textbf{Result:} This calculation yields a non-integer $\text{TF}_{\text{emb}}$ value, where words semantically central to the sentence receive a higher score.
\end{itemize}

\noindent\textbf{Embedding-aware Inverse Document Frequency ($\text{IDF}_{\text{emb}}$)} The Inverse Document Frequency (IDF) component measures a word's uniqueness across the entire document corpus. In the Embedding-level approach, the document frequency ($\text{DF}$) is calculated based on the similarity of a unique word's embedding to the embeddings of all tokens across all documents.

\begin{itemize}
    \item \textbf{Word-to-Document Similarity Matrix ($\text{Word2DocSim}$):} A $\text{Word2DocSim}$ matrix is calculated by taking the dot product between the unique word embeddings and the embeddings of all tokens (word pieces) in the corpus.
    \item \textbf{Document Frequency Accumulation:} The document frequency ($\text{DF}$) for a word $w$ is accumulated across the entire corpus by summing the squared $\text{Word2DocSim}$ values. This accumulation is performed sentence-by-sentence based on token availability.
    \item \textbf{IDF Calculation:} The final $\text{IDF}_{\text{emb}}$ is calculated using a log-normalization formula, where $N$ is the corpus size:
    \begin{equation}
    \text{IDF}_{\text{emb}}(w) = \log\left(\frac{N + 1}{\text{DF}(w) + 1}\right) + 1
    \end{equation}
\end{itemize}

\noindent\textbf{Final Weighting and Normalization} The $\text{TF}_{\text{emb}}$ is multiplied by the $\text{IDF}_{\text{emb}}$ to get the final raw TF-IDF weight for each non-special token. These weights are then normalized to ensure stability.

\clearpage
\section{Supplementary Validation Studies}
\label{appendix:F}
This section provides comprehensive validation of both the MECAT dataset and its evaluation metrics. We first validate data quality through automated pipeline filtering and two rounds of human evaluation (\Cref{sec:data_quality_validation}), and then verify metric reliability from three complementary perspectives: scoring bias, stability, and alignment with human judgments (\Cref{sec:metric_validation}).

\subsection{Data Quality Validation}
\label{sec:data_quality_validation}

The trustworthiness of MECAT annotations relies on a rigorous multi-stage pipeline. The expert models deployed (e.g., CED, Audio Flamingo 2) are architecturally independent with largely uncorrelated error distributions, making concurrent hallucination of the same content statistically unlikely. Through LLM-CoT domain conflict resolution and GLAP filtering, only ${\sim}$10\% of candidate samples from the CC-licensed source data survived the full pipeline for release. On this filtered set, two rounds of human evaluation were conducted to validate data quality from complementary perspectives: annotation scheme consistency with human judgment and annotation result quality at scale.

\subsubsection{Human Preference A/B Test}
\label{sec:human_ab_test}

A Human Preference A/B test ($N=150$ caption pairs spanning all domains) was conducted to verify that the annotation scheme produces outputs consistent with human judgment. Evaluators were instructed to select the better caption based on the stringent criterion: ``accuracy first, then level of detail.''

MECAT references (A) were compared against three opponent types (B) to probe specific quality aspects: Safe Captions (generic descriptions, testing discriminability); Wrong Captions (factually incorrect references, testing accuracy); and Human References (expert-written ground truth, establishing the quality ceiling).

The results are presented in \Cref{tab:human_pref_val}. MECAT references were strongly favored over both Safe and Wrong captions ($>94\%$ win rates), confirming the pipeline's effectiveness in mitigating vague or factually incorrect outputs. Crucially, the quality of MECAT references was found to be statistically on par with human-written references (56.9\% win rate, as the 95\% confidence interval spans 50\%).

\begin{table}[h]
\centering
\caption{Human Preference A/B Validation Results ($N=150$ pairs). Win rates indicate preference for the MECAT Reference (A).}
\label{tab:human_pref_val}
\begin{tabular}{l c c c}
\toprule
Opponent Type (B) & Size & Reference (A) Win Rate & 95\% CI (Wilson) \\
\midrule
Overall & 150 & 82.7\% & [75.8\%, 87.9\%] \\
\midrule
Safe Captions & 52 & 94.2\% & [84.4\%, 98.0\%] \\
Wrong Captions & 47 & 97.9\% & [88.9\%, 99.6\%] \\
Human References & 51 & \textbf{56.9\%} & [43.3\%, 69.5\%] \\
\bottomrule
\end{tabular}
\end{table}

\subsubsection{Expert Subjective Evaluation}
\label{sec:expert_eval}

To directly assess the quality of annotation results at scale, we recruited 20 audio-domain professionals to conduct an independent human audit. For each audio clip, the benchmark provides captions from multiple perspectives, including a global description and domain-specific descriptions from the speech, music, sound, and environment dimensions. Each participant listened to the original audio and judged whether these captions collectively captured the key details of the audio content, assigning a 1--5 Relevance (REL) score as defined in \Cref{tab:rel_scale}.

\begin{table}[h]
\centering
\caption{REL scoring scale definition for expert subjective evaluation.}
\label{tab:rel_scale}
\begin{tabular}{cl}
\toprule
Score & Interpretation \\
\midrule
5 & Perfect match -- captions precisely describe all audio details \\
4 & High match -- captions capture nearly all details; only minor omissions \\
3 & Basic match -- captions reflect the general intent of the audio \\
2 & Low match -- key details missing or incorrect \\
1 & No match -- captions are completely unrelated to the audio \\
\bottomrule
\end{tabular}
\end{table}

\noindent\textbf{Overall Results.}
Across 700 samples (100 per category), 85.3\% scored $\geq$3 (acceptable or better), 67.3\% achieved high accuracy (scores 4--5), and the mean REL was 3.81. The strict label error rate (REL\,=\,1) was bounded at only 3.4\% [95\% CI: 2.3\%--5.1\%]. Scores were significantly above the basic-match baseline of 3.0 (Wilcoxon $p < 0.001$; rank-biserial $r = 0.714$; Cohen's $d = 0.734$).

\noindent\textbf{Per-Category Results.}
Per-category results are presented in \Cref{tab:per_category_rel}. All seven categories passed the Wilcoxon signed-rank test ($p < 0.001$) against the 3.0 baseline after Bonferroni correction ($\alpha = 0.05/7$).

\begin{table}[h]
\centering
\caption{Per-category human audit results (H$_1$: REL $> 3.0$, 100 samples each).}
\label{tab:per_category_rel}
\begin{tabular}{llcccc}
\toprule
Category & Description & Mean REL & Acceptable ($\geq$ 3) & Cohen's $d$ & Rank-biserial $r$ \\
\midrule
00A & Sound Effects Only & 3.45 & 74.0\% & 0.365 & 0.422 \\
0M0 & Music Only & 3.77 & 80.0\% & 0.644 & 0.648 \\
0MA & Music + Sound Effects & 3.57 & 79.0\% & 0.453 & 0.511 \\
S00 & Speech Only & 4.10 & 95.0\% & 1.247 & 0.935 \\
S0A & Speech + Sound Effects & 3.89 & 88.0\% & 0.838 & 0.788 \\
SM0 & Speech + Music & 4.07 & 94.0\% & 1.250 & 0.915 \\
SMA & Speech + Music + Sound Effects & 3.85 & 87.0\% & 0.790 & 0.744 \\
\bottomrule
\end{tabular}
\end{table}

\noindent\textbf{Domain Variance.}
We observed meaningful scoring variance across audio domains: Speech (mean 3.98) $>$ Music (3.81) $>$ Sound (3.69). The difference between Speech and Sound domains is statistically significant (Mann--Whitney $U$, $p = 1.37 \times 10^{-3}$), with a small but measurable effect size (Cliff's $\delta = 0.125$; Cohen's $d = 0.267$). This quantifies the added difficulty of non-speech event alignment, particularly in categories where coarse metadata may not fully capture fine-grained acoustic details.

\begin{table}[h]
    \centering
    \caption{Model performance ($\text{Score}_{\text{Cap}}$) across different weighting scenarios. Kendall's $\tau = 0.92$ between Original and the other two settings.}
    \label{tab:weight_sensitivity}
    \begin{tabular}{lcccccc}
    \toprule
    \multirow{2}{*}{Model} & \multicolumn{2}{c}{Original (6:3:1)} & \multicolumn{2}{c}{Audio-Centric (2:4:4)} & \multicolumn{2}{c}{Equal (1:1:1)} \\
    \cmidrule(lr){2-3}\cmidrule(lr){4-5}\cmidrule(lr){6-7}
     & Score & Rank & Score & Rank & Score & Rank \\
    \midrule
    Gemini-3-Pro & 53.10 & 1 & 50.14 & 2 & 51.09 & 2 \\
    Qwen3-Omni-Flash-1201 & 52.90 & 2 & 50.39 & 1 & 51.20 & 1 \\
    Gemini-2.5-Flash & 51.60 & 3 & 49.08 & 3 & 49.82 & 3 \\
    Gemini-3-Flash & 51.10 & 4 & 47.80 & 5 & 48.91 & 5 \\
    Gemini-2.5-Pro & 50.60 & 5 & 48.34 & 4 & 49.04 & 4 \\
    Qwen2.5-Omni 7B & 42.60 & 6 & 41.88 & 7 & 42.21 & 6 \\
    Qwen2.5-Omni 3B & 42.50 & 7 & 41.94 & 6 & 42.14 & 7 \\
    Step-audio-2-mini & 41.50 & 8 & 39.54 & 8 & 40.16 & 8 \\
    Qwen3-Omni & 40.40 & 9 & 38.14 & 10 & 38.82 & 11 \\
    Audio Flamingo 3 & 40.40 & 10 & 38.16 & 9 & 38.94 & 9 \\
    MiMo-Audio-Instruct & 40.10 & 11 & 38.06 & 11 & 38.84 & 10 \\
    Baichuan-Omni & 35.60 & 12 & 33.01 & 12 & 33.91 & 12 \\
    Baichuan-Audio-Instruct & 33.70 & 13 & 31.39 & 14 & 32.30 & 14 \\
    Kimi-Audio-7B-Instruct & 32.80 & 14 & 32.34 & 13 & 32.59 & 13 \\
    Phi-4-Multimodal-Instruct & 30.00 & 15 & 28.88 & 15 & 29.29 & 15 \\
    \bottomrule
    \end{tabular}
    \end{table}
\subsection{Metric Validation}
\label{sec:metric_validation}

This subsection validates the reliability of the proposed evaluation metrics from three complementary perspectives: whether the scoring weights introduce systematic bias, whether the DATE metric produces stable scores across varying test conditions, and whether DATE aligns with human judgments.

\subsubsection{ScoreCap Weighting Sensitivity Analysis}
\label{sec:weight_sensitivity}

To verify that the heuristic weighting scheme for $\text{Score}_{\text{Cap}}$ does not bias model rankings toward speech-dominant models, we conducted a sensitivity analysis across 15 LALMs under three weighting scenarios: the Original data-driven weights (Speech/Music/Sound = 6:3:1), an Audio-Centric setting (2:4:4) that emphasizes non-speech domains, and an Equal setting (1:1:1). As shown in \Cref{tab:weight_sensitivity}, model rankings remain highly stable across all three scenarios, with Kendall's $\tau = 0.92$ between the Original and the other two settings. This confirms that top-performing models excel through comprehensive acoustic understanding rather than exploiting the speech-weighted distribution.

\subsubsection{DATE Bootstrap Stability Analysis}
\label{sec:date_bootstrap}

To assess the stability of DATE scores under varying test set sizes, we conducted a bootstrap stability analysis using Gemini-3.0-Pro outputs. Speech, Music, and Sound are each evaluated separately on Pure (single audio type) and Mixed (overlapping types) subsets with independent reference pools, yielding nine evaluation groups in total (including Long, Short, and Environment). For each group at each sampling ratio, 50 bootstrap iterations were performed, each fully re-computing DATE including the discrimination-rank component.

\noindent\textbf{Full-Scale Stability.}
\Cref{tab:bootstrap_full} summarizes bootstrap results at 100\% sampling. Eight out of nine groups exhibit a Coefficient of Variation (CV) below 1\%, with only Pure-Sound (CV\,=\,1.04\%, $N=848$) slightly exceeding this threshold due to its smaller pool size. The maximum deviation across all groups is bounded at $\pm 0.014$.

\begin{table}[h]
\centering
\caption{Full-data (100\%) bootstrap summary across nine evaluation groups (50 iterations each).}
\label{tab:bootstrap_full}
\begin{tabular}{lrcccc}
\toprule
Group & Size & DATE Mean & CV (\%) & 95\% CI & Max Dev \\
\midrule
Long & 20,052 & 0.6486 & 0.14 & [0.647, 0.650] & $\pm$0.002 \\
Short & 20,052 & 0.6581 & 0.16 & [0.656, 0.660] & $\pm$0.003 \\
Pure-Speech & 7,839 & 0.6049 & 0.29 & [0.602, 0.607] & $\pm$0.004 \\
Mixed-Speech & 8,593 & 0.6236 & 0.31 & [0.620, 0.627] & $\pm$0.004 \\
Pure-Music & 2,593 & 0.4967 & 0.77 & [0.490, 0.503] & $\pm$0.008 \\
Mixed-Music & 8,593 & 0.3988 & 0.46 & [0.395, 0.402] & $\pm$0.004 \\
Pure-Sound & 848 & 0.5520 & 1.04 & [0.542, 0.565] & $\pm$0.014 \\
Mixed-Sound & 8,593 & 0.2992 & 0.94 & [0.294, 0.304] & $\pm$0.007 \\
Environment & 20,052 & 0.2613 & 0.58 & [0.259, 0.264] & $\pm$0.003 \\
\bottomrule
\end{tabular}
\end{table}

\noindent\textbf{Convergence Across Subset Sizes.}
\Cref{tab:bootstrap_cv} reports the CV across different sampling ratios. CV drops below 3\% at 5\% sampling for seven out of nine groups. Even for the smallest group (Pure-Sound, $N=848$), CV reaches below 6\% at 5\% sampling. At $\geq$10\% sampling, all groups except Pure-Sound and Pure-Music fall below 2.5\%.

\begin{table}[h]
\centering
\caption{CV (\%) across subset sizes. CV@$x$\% denotes CV from 50 bootstrap iterations each drawing $x$\% of full data with replacement.}
\label{tab:bootstrap_cv}
\begin{tabular}{lcccccc}
\toprule
Group & CV@1\% & CV@5\% & CV@10\% & CV@20\% & CV@50\% & CV@100\% \\
\midrule
Long & 1.79 & 0.61 & 0.54 & 0.32 & 0.22 & 0.14 \\
Short & 1.60 & 0.58 & 0.55 & 0.36 & 0.22 & 0.16 \\
Pure-Speech & 3.14 & 1.37 & 0.92 & 0.68 & 0.40 & 0.29 \\
Mixed-Speech & 3.53 & 1.22 & 0.88 & 0.67 & 0.47 & 0.31 \\
Pure-Music & 10.65 & 4.10 & 3.22 & 2.29 & 1.32 & 0.77 \\
Mixed-Music & 5.86 & 2.69 & 2.28 & 1.37 & 0.67 & 0.46 \\
Pure-Sound & 18.62 & 5.85 & 4.61 & 3.22 & 1.68 & 1.04 \\
Mixed-Sound & 9.60 & 4.36 & 3.11 & 2.77 & 1.28 & 0.94 \\
Environment & 5.60 & 2.92 & 1.90 & 1.32 & 0.98 & 0.58 \\
\bottomrule
\end{tabular}
\end{table}

These results confirm that DATE produces stable, scale-invariant scores across all evaluation groups, supporting its reliability as an evaluation metric for fine-grained audio captioning.

\subsubsection{Alignment of DATE with Human Judgments}
\label{sec:date_human_alignment}

The alignment of the DATE metric with human preferences was assessed using the same 150 A/B caption pairs from \Cref{sec:human_ab_test}, where listener choices established the human ``gold standard''.

For this validation, the human-preferred caption and the non-preferred caption from each A/B pair were separately treated as a candidate hypothesis. The DATE score was then computed for each candidate against the audio segment's original ground truth reference.

A comparison of mean DATE scores, presented in \Cref{tab:date_human_alignment}, reveals a substantial difference: captions selected by human evaluators received substantially higher mean DATE scores (90.9) compared to non-preferred captions (49.3). This large margin ($\Delta \approx 41.6$) robustly demonstrates that DATE is highly correlated with human preference regarding accuracy and detail, thereby validating its utility as a fine-grained evaluation tool.

\begin{table}[h]
\centering
\caption{Alignment between DATE Score and Human Preference ($N=150$ A/B pairs).}
\label{tab:date_human_alignment}
\begin{tabular}{l c}
\toprule
Caption Group & DATE \\
\midrule
Human-Preferred Captions & \textbf{90.9} \\
Non-Preferred Captions & 49.3 \\
\bottomrule
\end{tabular}
\end{table}

\clearpage
\section{LLM-as-Judge Prompts in evaluation}
\label{appendix:G}
This section provides the prompt template required for LLM-as-Judge method. The evaluation \textit{tasks} primarily include audio captioning and audio question-answering. In the template, the \textit{description}, \textit{subtask}, and \textit{scoring\_aspects} parameters can be referenced from the corresponding columns in the task table above, while \textit{ref\_texts} represents the samples to be evaluated.

\subsection{\textnormal{Evaluation Prompt Template}}

You are tasked with evaluating if a set of candidate \{tasks\} responses accurately addresses the same audio as a reference set of answers. You will focus on the \{description\} for the subtask '\{subtask\}'.

\textbf{Evaluation Steps:}
\begin{enumerate}[label=\alph*), leftmargin=3em]
\item First, carefully compare the candidate answers with the reference answers
\item Assess the accuracy and precision of how the audio characteristics are captured in the responses, then provide a 0-10 fine-grained score:
   \begin{itemize}[leftmargin=1em]
   \item 10 = perfect match with the reference content
   \item 0 = completely wrong
   \end{itemize}
\item Provide detailed scoring reasoning, explaining why you gave this score
\end{enumerate}

\textbf{Scoring Aspects:} \{scoring\_aspects\}

\textbf{Score rubric (0-10 Scale):}
\begin{itemize}[leftmargin=3em]
\item points 9-10: Excellent - Highly accurate, comprehensive, well-expressed
\item points 8:~~~~ Very Good - Accurate with minor gaps, clear expression
\item points 7:~~~~ Good - Mostly accurate, some missing details
\item points 6:~~~~ Acceptable - Basic accuracy, meets minimum HIGH standard
\item points 4-5:~ Below Standard - Some correct elements but major issues
\item points 2-3:~ Poor - Limited accuracy, significant problems
\item points 0-1:~ Very Poor - Major errors or completely incorrect
\end{itemize}

\textbf{You need to evaluate the following sample:} \{ref\_texts\}

\textbf{Please return JSON-formatted evaluation results for the sample.}

\quad Return format (strict JSON array):

\quad\quad sample\_id: sample ID

\quad\quad subtask: subtask\_name

\quad\quad fine\_score: $<$numerical value 0-10$>$

\quad\quad reasoning: detailed scoring rationale, including 
comparative analysis with reference answers

\begin{table}[htbp]
    \centering
    \caption{Category, subcategory, descriptions and scoring aspects of captioning evaluation with LLM-as-Judge method}
    \resizebox{\linewidth}{!}{%
    \begin{tabular}{llp{11cm}}
    \toprule
    Category & Subcategory & Description \& Scoring Aspects \\
    \midrule
    \multirow{11}{*}{\makecell[c]{Systemic}} & \multirow{5}{*}{\makecell[c]{Short}} & \textbf{Description:} quality of short audio descriptions \\
    & & \textbf{Scoring Aspects:} \\
    & & \quad a) accuracy of core content capture (most important) \\
    & & \quad b) conciseness and completeness of expression \\
    & & \quad c) semantic consistency with reference descriptions \\
    \cmidrule(lr){2-3}
    & \multirow{5}{*}{\makecell[c]{Long}} & \textbf{Description:} quality of detailed audio descriptions \\
    & & \textbf{Scoring Aspects:} \\
    & & \quad a) comprehensiveness and richness of description details \\
    & & \quad b) accuracy of detailed descriptions \\
    & & \quad c) logical structure and expression coherence \\
    \midrule
    \multirow{14}{*}{\makecell[c]{Content-\\Specific}} & \multirow{5}{*}{\makecell[c]{Speech}} & \textbf{Description:} accuracy of speech content recognition \\
    & & \textbf{Scoring Aspects:} \\
    & & \quad a) accuracy rate of speech content recognition \\
    & & \quad b) accurate description of speaker characteristics (gender, accent, etc.) \\
    & & \quad c) description of speech quality and environment \\
    \cmidrule(lr){2-3}
    & \multirow{5}{*}{\makecell[c]{Music}} & \textbf{Description:} quality of music content description \\
    & & \textbf{Scoring Aspects:} \\
    & & \quad a) accuracy of music type, style, and rhythm identification \\
    & & \quad b) identification of instruments and musical elements \\
    & & \quad c) description of musical emotion and atmosphere \\
    \cmidrule(lr){2-3}
    & \multirow{5}{*}{\makecell[c]{Sound}} & \textbf{Description:} accuracy of sound event identification \\
    & & \textbf{Scoring Aspects:} \\
    & & \quad a) accurate identification and classification of sound sources \\
    & & \quad b) description of sound occurrence timing and duration \\
    & & \quad c) description of sound intensity, pitch and other characteristics \\
    \midrule
    \multirow{5}{*}{\makecell[c]{Content-\\Unrelated}} & \multirow{5}{*}{\makecell[c]{Environment}} & \textbf{Description:} accuracy of environment and recording quality description \\
    & & \textbf{Scoring Aspects:} \\
    & & \quad a) identification of recording environment (indoor/outdoor, space size, etc.) \\
    & & \quad b) assessment of audio technical quality (distortion, noise, etc.) \\
    & & \quad c) description of environmental atmosphere and background characteristics \\
    \bottomrule
    \end{tabular}
    }
    \end{table}
\begin{table}[htbp]
    \caption{Category, subcategory, descriptions and scoring aspects of question-answering evaluation with LLM-as-Judge method}
    \label{tab:tasks_subtasks_qa}
    \resizebox{\linewidth}{!}{%
    \begin{tabular}{llp{11cm}}
    \toprule
    Category & Subcategory & Description \& Scoring Aspects \\
    \midrule
    \multirow{5}{*}{\makecell[c]{Perception}} & \multirow{5}{*}{\makecell[l]{Direct\\ Perception}} & \textbf{Description:} accuracy of direct audio content identification \\
    & & \textbf{Scoring Aspects:} \\
    & & \quad a) correct identification of primary audio elements (most important) \\
    & & \quad b) accurate detection of presence/absence of specific sounds \\
    & & \quad c) precise recognition of obvious audio features and events \\
    \midrule
    \multirow{10}{*}{\makecell[c]{Analysis}} & \multirow{5}{*}{\makecell[l]{Sound\\ Characteristics}} & \textbf{Description:} quality of sound property analysis \\
    & & \textbf{Scoring Aspects:} \\
    & & \quad a) accurate description of sound attributes (pitch, volume, timbre, etc.) \\
    & & \quad b) correct identification of sound sources and their properties \\
    & & \quad c) precise characterization of audio dynamics and patterns \\
    \cmidrule(lr){2-3}
    & \multirow{5}{*}{\makecell[l]{Quality\\ Assessment}} & \textbf{Description:} accuracy of audio quality evaluation \\
    & & \textbf{Scoring Aspects:} \\
    & & \quad a) correct assessment of technical audio quality (clarity, distortion, etc.) \\
    & & \quad b) accurate evaluation of recording conditions and fidelity \\
    & & \quad c) appropriate judgment of audio production quality \\
    \midrule
    \multirow{15}{*}{\makecell[c]{Reasoning}} & \multirow{5}{*}{\makecell[l]{Environment\\ Reasoning}} & \textbf{Description:} quality of environmental context inference \\
    & & \textbf{Scoring Aspects:} \\
    & & \quad a) accurate inference of recording location and setting \\
    & & \quad b) correct identification of spatial and acoustic properties \\
    & & \quad c) logical deduction of environmental factors affecting audio \\
    \cmidrule(lr){2-3}
    & \multirow{5}{*}{\makecell[l]{Inference\\ Judgment}} & \textbf{Description:} accuracy of complex audio analysis and reasoning \\
    & & \textbf{Scoring Aspects:} \\
    & & \quad a) correct interpretation of implicit audio information \\
    & & \quad b) accurate temporal reasoning and sequence understanding \\
    & & \quad c) logical inference of causality and relationships between audio elements \\
    \cmidrule(lr){2-3}
    & \multirow{5}{*}{\makecell[l]{Application\\ Context}} & \textbf{Description:} relevance and appropriateness of contextual understanding \\
    & & \textbf{Scoring Aspects:} \\
    & & \quad a) accurate understanding of audio's intended purpose or context \\
    & & \quad b) appropriate application of domain-specific knowledge \\
    & & \quad c) correct interpretation of cultural, social, or professional context \\
    \bottomrule
    \end{tabular}
    }
    \end{table}

\clearpage
\section{Validation of LLM-as-Judge as a reference metric}
\label{appendix:H}
To validate the effectiveness of LLM-as-Judge as a reference metric, we assessed its performance on three distinct sets of responses with varying quality levels: Right (detailed and accurate rephrasings of the ground-truth reference), Safe (generic, vague descriptions, e.g., "A man is speaking" for all speech-only audio), and Wrong (factually incorrect references randomly selected from other samples). As shown in following table, our analysis confirms that LLM-as-Judge method serves as a reliable evaluator. It successfully distinguishes between the quality tiers, with mean scores consistently following the expected Right~$>$~Safe~$>$~Wrong order for both captioning and QA tasks. Furthermore, its inter-rater reliability, measured by Fleiss' Kappa ($\kappa$), is substantial for QA ($\kappa=0.73$) and moderate for captioning ($\kappa=0.43$). However, the significant practical limitations of LLM-as-Judge method---including high computational cost, slow speed, and sensitivity to prompt engineering---motivate our development of the DATE metric as an efficient and scalable alternative.
\begin{table}[h]
    \centering
    \caption{Reliability of LLM-as-Judge (M.J.). Evaluation on three sets: \textbf{Right'} (rephrased references), \textbf{Safe'} (generic/vague domain responses), and \textbf{Wrong'} (incorrect random references). $\kappa$ denotes Fleiss' Kappa for inter-rater reliability; Mean' is the average M.J. score.}
    \label{tab:llm_judge_reliability}
    \begin{tabular}{@{}l@{\hspace{0.2em}}c@{\hspace{0.2em}}c@{\hspace{0.2em}}c@{\hspace{0.2em}}c@{}}
    \toprule
    \multirow{2}{*}{Type} & \multicolumn{2}{c}{Mean} & \multicolumn{2}{c}{Fleiss' Kappa ($\kappa$)} \\ 
    \cmidrule(lr){2-3}\cmidrule{4-5}
                          & Caption & QA & Caption & QA \\ 
    \midrule
    Right & 0.78 & 0.97 & 0.68 & 0.74 \\
    Safe & 0.24 & - & 0.17 & - \\
    Wrong & 0.13 & 0.12 & 0.45 & 0.72 \\ 
    \midrule
    Overall & - & - & 0.43 & 0.73 \\ 
    \bottomrule
    \end{tabular}
\end{table}

\clearpage
\section{Task-specific prompts for LALM in MECAT tasks}
\label{appendix:I}
This section details the prompt strategies employed for Large Audio-Language Models (LALMs) during the MECAT-Caption evaluation. For specific Audio-focused LALMs that require specialized instruction formats---namely Audio-Flamingo 2~\citep{ghosh2025audio} and Kimi-Audio~\citep{ding2025kimi}---the exact prompt templates are provided in \Cref{tab:audio_caption_prompts}.  For the remaining LALMs, the prompts are standardized as shown in \Cref{tab:qwen_omni_prompts}. This category encompasses a diverse range of architectures, including other Audio-focused models such as MiMo-Audio~\citep{zhang2025mimo}, Step-Audio-2-mini~\citep{wu2025stepaudio2}, and Baichuan-Audio~\citep{li2025baichuan}; Omni LALMs represented by the Qwen-Omni series~\citep{xu2025qwen2,xu2025qwen3,qwen2025flash1201} and Baichuan-Omni~\citep{li2024baichuan}; Multimodal LALMs specifically Phi-4-Multimodal~\citep{phi42025}; and the state-of-the-art Gemini series~\citep{comanici2025gemini,google2025gemini3}. Regarding system configurations, the system prompt was explicitly set to ``\textit{You are a helpful assistant}'' for the Qwen2.5-Omni models, while the default system prompts were retained for all other architectures.

Regarding the MECAT-QA task, the prompt for each sample consists solely of the corresponding question, without additional task-specific templates.

\begin{table}[htbp]
    \centering
    \caption{Prompts for Audio-Flamingo2 and Kimi-Audio models in caption task}
    \label{tab:audio_caption_prompts}
    \begin{tabular}{llp{8cm}}
    \toprule
    Category & Subcategory & Prompt \\
    \midrule
    \multirow{2}{*}{Systematic} & Short & Provide a caption for this audio within 15 words \\
                               & Long & Provide a caption for this audio within 1-2 sentences \\
    \midrule
    \multirow{4}{*}{Content-Specific} & Speech & Provide a caption for the speech content in this audio \\
                                    & Music  & Provide a caption for the music content in this audio \\
                                    & Sound  & Provide a caption for general sound excluding speech and music \\
    \midrule
    \multirow{2}{*}{Content-Unrelated} & \multirow{2}{*}{Environment} & Provide a caption for quality or acoustic environment for this audio \\
    \bottomrule
    \end{tabular}
\end{table}

\begin{table*}[ht]
    \centering
    \caption{Prompts for remaining models in caption task}
    \label{tab:qwen_omni_prompts}
    \begin{tabular}{llp{8cm}}
    \toprule
    Category & Subcategory & Prompt \\
    \midrule
    \multirow{4}{*}{Systematic} & \multirow{2}{*}{Short} & Listen to the audio and provide a caption for this audio within 15 words \\
                               & \multirow{2}{*}{Long} & Listen to this audio and provide a caption for this audio within 1-2 sentences \\
    \midrule
    \multirow{6}{*}{Content-Specific} & \multirow{2}{*}{Speech} & Listen to the audio and provide a caption describing the speech content in this audio \\
                                    & \multirow{2}{*}{Music} & Listen to the audio and provide a caption for the music content in this audio \\
                                    & \multirow{2}{*}{Sound} & Listen to the audio and provide a general sound excluding speech and music \\
    \midrule
    \multirow{2}{*}{Content-Unrelated} & \multirow{2}{*}{Environment} & Listen to this audio and provide a caption for quality or acoustic environment for this audio \\
    \bottomrule
    \end{tabular}
    \end{table*}

\clearpage
\section{Fine-grained Analysis of Speech Captioning}
\label{appendix:J}
This section presents a detailed performance of all models on the pure speech subset of MECAT-Caption task.  \Cref{tab:similarity_date_discri} presents the results across three dimensions: similarity, discriminability, and DATE. 

\begin{table}[h]
\centering
\caption{Comparison of different models. $^\dagger$ indicates that its previous version (Audio Flamingo 2) was explicitly used in the data construction process.}
\label{tab:similarity_date_discri}
\newcommand{\grank}[1]{\textcolor{gray}{#1}}
\begin{tabular}{@{}llcc cc cc@{}}
\toprule
\multirow{2}{*}{Type} & \multirow{2}{*}{Model} & \multicolumn{2}{c}{Similarity} & \multicolumn{2}{c}{Discriminability} & \multicolumn{2}{c}{DATE} \\
\cmidrule(lr){3-4} \cmidrule(lr){5-6} \cmidrule(lr){7-8}
 & & Score & \grank{Rank} & Score & \grank{Rank} & Score & \grank{Rank} \\ 
\midrule
\multirow{2}{*}{\makecell[l]{Caption\\-Only}} 
 & Pengi & 26.6 & \grank{15} & 27.8 & \grank{16} & 27.2 & \grank{16} \\
 & EnClap & 28.7 & \grank{14} & 31.9 & \grank{15} & 30.2 & \grank{14} \\ 
\midrule
\multirow{14}{*}{LALM} 
 & Phi-4-Multimodal-Instruct & 26.6 & \grank{16} & 27.2 & \grank{17} & 26.9 & \grank{17} \\
 & Kimi-Audio-7B-Instruct & 25.6 & \grank{17} & 36.2 & \grank{14} & 30.0 & \grank{15} \\
 & Baichuan-Audio-Instruct & 37.2 & \grank{10} & 60.3 & \grank{7} & 46.0 & \grank{8} \\
 & Audio Flamingo 3$^\dagger$ & 46.6 & \grank{4} & 52.3 & \grank{10} & 49.3 & \grank{7} \\
 & MiMo-Audio-Instruct & 42.5 & \grank{7} & 49.7 & \grank{11} & 45.8 & \grank{9} \\
 & Step-Audio-2-mini & 36.6 & \grank{11} & 55.8 & \grank{9} & 44.2 & \grank{10} \\
 & Baichuan-Omni & 34.9 & \grank{13} & 57.7 & \grank{8} & 43.5 & \grank{11} \\
 & Qwen2.5-Omni 3B & 37.3 & \grank{9} & 49.4 & \grank{12} & 42.5 & \grank{12} \\
 & Qwen2.5-Omni 7B & 35.3 & \grank{12} & 45.9 & \grank{13} & 39.9 & \grank{13} \\
 & Qwen3-Omni & 41.0 & \grank{8} & 64.7 & \grank{6} & 50.2 & \grank{6} \\
 \cmidrule(lr){2-8}
 & Qwen3-Omni-Flash-1201 & \underline{47.7} & \grank{2} & \textbf{80.8} & \grank{1} & 59.2 & \grank{3}\\ 
 & Gemini-2.5-Flash & 45.8 & \grank{5} & 77.2 & \grank{5} & 57.5 & \grank{4} \\
 & Gemini-2.5-Pro & 44.3 & \grank{6} & 78.4 & \grank{4} & 56.6 & \grank{5} \\ 
 & Gemini-3-Flash & 47.5 & \grank{3} & 79.3 & \grank{3} & \underline{59.4} & \grank{2} \\
 & Gemini-3-Pro & \textbf{48.8} & \grank{1} & \underline{79.6} & \grank{2} & \textbf{60.5} & \grank{1} \\
\bottomrule
\end{tabular}
\end{table}

\clearpage
\section{Model performance of similarity on MECAT Tasks}
\label{appendix:K}
This section presents the complete Similarity scores for all models evaluated on MECAT, serving as a comparative reference for the DATE metrics reported in the main text (see Tables~\ref{tab:bleu1_results} and~\ref{tab:qa_results}).

\begin{table}[ht]
\centering
\caption{Model performance (Similarity \%) on MECAT-Caption. \textbf{Bold} indicates the best performance, and \underline{underline} indicates the second best. $^\dagger$ indicates that its previous version (Audio Flamingo 2) was explicitly used in the data construction process.}
\label{tab:bleu1_results_similarity}
\resizebox{\textwidth}{!}{%
\begin{tabular}{@{}llc*{3}{c}*{6}{c}c@{}}
\toprule
\multirow{4}{*}{Type} & 
\multirow{4}{*}{Model} & 
\multicolumn{2}{c}{Systemic} & 
\multicolumn{6}{c}{Content-Specific} & 
\multirow{3}{*}{\makecell[c]{Content\\Unrelated}} & 
\multirow{4}{*}{Score$_\text{Cap}$} \\
\cmidrule(lr){3-4}\cmidrule(lr){5-10} 
 & & 
\multirow{2}{*}{\makecell[l]{Long}} & \multirow{2}{*}{\makecell[l]{Short}} & 
\multicolumn{2}{c}{Speech} & 
\multicolumn{2}{c}{Music} & 
\multicolumn{2}{c}{Sound} & \\
\cmidrule(lr){5-10} \cmidrule(lr){11-11} & & & & 
Pure & Mixed & 
Pure & Mixed & 
Pure & Mixed & 
\multicolumn{1}{c}{Env} & 
\multicolumn{1}{c}{} & \\
\midrule
\multirow{2}{*}{\makecell[c]{Caption\\-Only}} 
 & Pengi & 37.5 & 41.0 & 26.6 & 29.2 & 39.6 & 11.8 & 35.4 & 16.2 & 17.8 & 29.5 \\
 & EnClap & 40.5 & 45.0 & 28.7 & 29.5 & 39.3 & 15.0 & 41.2 & 17.3 & 17.9 & 31.6 \\
\midrule
\multirow{15}{*}{LALM} 
 & Phi-4-Multimodal-Instruct & 45.4 & 40.3 & 26.6 & 31.7 & 41.5 & 26.2 & 29.5 & 25.7 & 37.3 & 37.4 \\
 & Kimi-Audio-7B-Instruct & 40.8 & 45.7 & 25.6 & 27.1 & 39.5 & 16.2 & 35.8 & 19.4 & 16.7 & 30.8 \\
 & Baichuan-Audio-Instruct & 33.0 & 28.2 & 37.2 & 35.0 & 36.4 & 24.7 & 45.0 & 29.9 & 47.1 & 36.1 \\
 & Baichuan-Omni & 39.2 & 42.5 & 34.9 & 35.4 & 41.0 & 13.2 & 40.0 & 32.3 & 29.4 & 35.0 \\
 & MiMo-Audio-Instruct & 49.9 & 49.4 & 42.5 & 43.5 & 47.5 & 19.9 & 44.5 & 27.6 & 27.2 & 41.2 \\
 & Audio Flamingo 3$^\dagger$ & 49.6 & 49.6 & 46.6 & 47.5 & 50.6 & 26.4 & 44.6 & 28.3 & 31.7 & 43.5 \\
 & Qwen3-Omni & 38.2 & 33.6 & 34.1 & 34.5 & 49.0 & 34.1 & 41.4 & 20.8 & 40.2 & 37.4 \\
 & Step-Audio-2-mini & 44.1 & 47.8 & 36.6 & 37.3 & 45.9 & 36.0 & 36.4 & 24.9 & 41.4 & 41.2 \\
 & Qwen2.5-Omni 3B & 48.3 & 45.3 & 37.3 & 37.5 & 50.7 & 34.7 & 46.6 & 34.1 & 47.8 & 44.1 \\
 & Qwen2.5-Omni 7B & 52.7 & 46.2 & 35.3 & 37.5 & 39.2 & 33.1 & 45.2 & 32.1 & 41.0 & 43.4 \\
 \cmidrule(lr){2-12}
 & Qwen3-Omni-Flash-1201 & \underline{53.6} & 50.4 & 46.7 & 47.6 & \textbf{61.8} & 38.2 & 50.4 & \textbf{36.5} & \textbf{52.7} & 50.7 \\
 & Gemini-2.5-Flash & \textbf{56.1} & \textbf{53.5} & 45.8 & 46.6 & \underline{59.1} & \underline{44.3} & \textbf{50.7} & \underline{36.4} & 48.9 & \underline{51.0} \\
 & Gemini-2.5-Pro & 50.8 & 49.9 & 44.3 & 45.7 & 58.5 & \textbf{44.6} & 49.6 & 35.0 & \underline{51.9} & 49.3 \\
 & Gemini-3-Flash & 53.1 & 50.8 & \underline{47.5} & \underline{49.2} & 53.6 & 42.6 & 49.8 & 36.0 & 50.8 & 50.3 \\
 & Gemini-3-Pro & 53.2 & \underline{53.3} & \textbf{48.8} & \textbf{50.7} & 58.0 & 43.1 & \underline{50.5} & 34.9 & 49.4 & \textbf{53.1} \\
\bottomrule
\end{tabular}%
}
\end{table}

\begin{table}[ht]
\centering
\caption{Model Performance (Similarity \%) on MECAT-QA. \textbf{Bold} indicates the best performance, and \underline{underline} indicates the second best. $^\dagger$ indicates that its previous version (Audio Flamingo 2) was explicitly used in the data construction process.}
\label{tab:qa_results_similarity}
\resizebox{\textwidth}{!}{%
\begin{tabular}{@{}lc*{2}{c}*{3}{c}c@{}}
\toprule
\multirow{2}{*}{Model} & 
\multicolumn{1}{c}{Perception} & 
\multicolumn{2}{c}{Analysis} & 
\multicolumn{3}{c}{Reasoning} & 
\multirow{2}{*}{Score$_{\text{QA}}$} \\
\cmidrule(lr){2-2}\cmidrule(lr){3-4}\cmidrule(lr){5-7}
 & 
\makecell[c]{Direct\\Perception} & 
\makecell[c]{Sound\\Characteristics} & 
\makecell[c]{Quality\\Assessment} & 
\makecell[c]{Environment\\Reasoning} & 
\makecell[c]{Inference \&\\Judgment} & 
\makecell[c]{Application\\Context} & 
 \\
\midrule
 Kimi-Audio-7B-Instruct & 37.5 & 32.5 & 19.2 & 37.5 & 38.8 & 33.8 & 33.2 \\
 Baichuan-Audio-Instruct & 35.2 & 36.6 & 36.0 & 38.1 & 39.5 & 39.6 & 37.5 \\
 Baichuan-Omni & 36.8 & 36.1 & 35.4 & 39.1 & 38.5 & 39.4 & 37.6 \\
 Phi-4-Multimodal-Instruct & 41.2 & 37.6 & 36.6 & 40.3 & 39.0 & 40.1 & 39.1 \\
 MiMo-Audio-Instruct & \underline{50.9} & 40.5 & 27.0 & 40.7 & 41.9 & 38.5 & 39.9 \\
 Step-Audio-2-mini & 48.6 & 44.6 & 39.1 & 38.2 & 38.7 & 39.3 & 41.4 \\
 Audio Flamingo 3$^\dagger$ & 46.0 & 41.4 & 38.6 & 43.5 & 43.2 & 40.9 & 42.3 \\
 Qwen2.5-Omni 3B & 47.2 & 43.8 & 39.7 & 43.2 & 41.0 & 41.9 & 42.8 \\
 Qwen2.5-Omni 7B & 49.7 & 43.8 & \underline{40.5} & 44.1 & 42.5 & 41.9 & 43.8 \\
 Qwen3-Omni & \textbf{52.3} & 44.8 & \textbf{41.2} & 45.2 & 44.7 & 45.2 & \underline{45.6} \\
 \cmidrule(lr){1-8}
 Qwen3-Omni-Flash-1201 & 41.3 & 38.5 & 35.8 & 44.4 & \underline{45.9} & 45.6 & 41.9 \\
 Gemini-2.5-Flash & 47.9 & \textbf{46.1} & 39.7 & \underline{46.2} & \textbf{47.1} & \textbf{47.9} & \textbf{45.8} \\ 
 Gemini-2.5-Pro & 47.4 & \underline{45.2} & 39.0 & \textbf{46.9} & 45.7 & \underline{46.3} & 45.1 \\
 Gemini-3-Flash & 46.0 & 42.3 & 34.0 & 44.7 & 45.4 & 46.1 & 43.1 \\
 Gemini-3-Pro & 47.4 & 37.8 & 26.1 & 41.2 & 42.3 & 41.9 & 39.5 \\
\bottomrule
\end{tabular}%
}
\end{table}

\clearpage
\section{Analysis of Speech Hallucination in Silent Segments}
\label{appendix:L}
This section qualitatively evaluates the grounding capabilities of various models when presented with audio segments containing no discernible speech. We specifically analyze the "speech" output field for two representative silent cases (C1 and C2) to detect potential hallucinations in low-signal environments. 
\begin{table}[h]
\centering
\caption{Speech caption generated by different models for two silence segments in MECAT. $^\dagger$ indicates that its previous version (Audio Flamingo 2) was explicitly used in the data construction process.}
\label{tab:silence_results_appendix}


\resizebox{\textwidth}{!}{%
    \begin{tabularx}{1.15\textwidth}{m{2.2cm} m{3.8cm} X} 
    \toprule
    Type & Model & Output \\ 
    \midrule
    
    \multirow{4}{*}{Caption-Only} 
     & Pengi & \textbf{C1:} a person is playing a guitar. \newline 
                        \textbf{C2:} someone is opening and closing a door. \\ 
     \cmidrule(l){3-3}
     & EnClap & \textbf{C1:} A low frequency of a person talking in the distance \newline 
                         \textbf{C2:} A helicopter flying in the distance followed by a man speaking \\ 
    \midrule
    
    \multirow{46}{*}{LALM} 
     & Phi-4-Multimodal-Instruct & \textbf{C1:} Audio caption: A rumble is heard. \newline 
                                            \textbf{C2:} Audio caption: A man is speaking with music in the background. \\ 
     \cmidrule(l){3-3}
     
     & Kimi-Audio-7B-Instruct & \textbf{C1:} a machine is running \newline 
                                         \textbf{C2:} a person is playing a drum set with a bass drum and a snare drum. \\ 
     \cmidrule(l){3-3}
     
     & Baichuan-Audio-Instruct & \textbf{C1:} The speaker is talking about how they are going through their day... [Truncated] \newline 
                                          \textbf{C2:} The speaker is asking for help with their website... [Truncated] \\ 
     \cmidrule(l){3-3}
     
     
     & Baichuan-Omni & \textbf{C1:} A woman is speaking with background noise. \newline 
                                \textbf{C2:} Someone is speaking with authority, likely giving instructions or delivering an important message. \\ 
     \cmidrule(l){3-3}
     
     & MiMo-Audio-Instruct & \textbf{C1:} A male voice, speaking in a neutral tone... says 'Thank you' in English... \newline 
                                      \textbf{C2:} The audio contains a sequence of sound effects... with no speech. \\ 
     \cmidrule(l){3-3}
     
     & Audio Flamingo 3$^\dagger$ & \textbf{C1:} A male voice says 'Thank you' amidst the sound of a waterfall. \newline 
                                   \textbf{C2:} A female voice says 'Thank you' in a neutral tone. \\ 
     \cmidrule(l){3-3}
     
     & Qwen3-Omni & \textbf{C1:} A person is speaking. \newline 
                             \textbf{C2:} The audio contains only a single, sustained, low-pitched electronic tone. \\ 
     \cmidrule(l){3-3}
     
     & Step-audio-2-mini & \textbf{C1:} The speech content is "Oh no, I'm sorry." \newline 
                                    \textbf{C2:} There is no speech in this audio. \\ 
     \cmidrule(l){3-3}
     
     & Qwen2.5-Omni 3B & \textbf{C1:} The audio contains a speech saying 'I'm gonna be a daddy'. \newline 
                                  \textbf{C2:} The audio contains a speech segment where a male voice says 'you' in a neutral tone. \\ 
     \cmidrule(l){3-3}
     
     & Qwen2.5-Omni 7B & \textbf{C1:} The audio contains a speech segment where someone is saying 'I'm going to go ahead and do that.' \newline 
                                  \textbf{C2:} The audio contains a speech segment in which the speaker says 'you'. \\ 
     \cmidrule(l){2-3}

     & Qwen3-Omni-Flash-1201 & \textbf{C1:} No speech detected in the audio.
     \newline
                               \textbf{C2:} There is no speech in this audio. \\
    \cmidrule(l){3-3}
     
     & Gemini-2.5-Flash & \textbf{C1:} No discernible speech is present in this audio. \newline 
                                   \textbf{C2:} No speech detected. \\ 
     \cmidrule(l){3-3}
     
     & Gemini-2.5-Pro & \textbf{C1:} There is no speech in this audio. \newline 
                                 \textbf{C2:} There is no speech in this audio. \\ 
    \cmidrule(l){3-3}

    & Gemini-3-Flash & \textbf{C1:} A man provides a continuous spoken delivery, though the clarity is slightly affected by room acoustics. 
    \newline
                        \textbf{C2:} A male voice counts 'uno, dos, tres' in Spanish before the guitar playing begins. \\
    \cmidrule(l){3-3}            
    & Gemini-3-Pro   & \textbf{C1:} The audio does not contain any speech or vocalizations. 
    \newline
                    \textbf{C2:} A male voice speaks with an urgent and hushed tone, delivering a dramatic monologue. \\
    \bottomrule
    \end{tabularx}%
}
\end{table}
\clearpage

\end{document}